\documentclass[trackchanges,twocolumn,tighten]{aastex62} 

\usepackage{amsmath}

\usepackage{longtable}
\usepackage{soul}
\usepackage{CJK}

\newcommand{\code}[1]{\texttt{#1}}

\newcommand{\mesa}{\code{MESA}}
\newcommand{\MESA}{\mesa}

\newcommand{\Athena}{\code{Athena++}}
\newcommand{\COBOLD}{\code{CO}$^\texttt{5}$\code{BOLD}}
\newcommand{\vavg}[1]{\langle{#1}\rangle}
\newcommand{\Msun}{M_\odot}
\newcommand{\Rsun}{R_\odot}
\newcommand{\Lsun}{L_\odot}
\newcommand{\vc}{v_c}
\renewcommand{\vr}{v_r}
\newcommand{\vtheta}{v_\theta}
\newcommand{\vphi}{v_\phi}
\newcommand{\Prad}{P_\mathrm{rad}}
\newcommand{\Pgas}{P_\mathrm{gas}}
\newcommand{\Pturb}{P_\mathrm{turb}}
\newcommand{\Ptherm}{P_\mathrm{therm}}
\newcommand{\Ptot}{P_\mathrm{tot}}
\newcommand{\taut}{\tau_\mathrm{crit}}
\newcommand{\ah}{\alpha_\mathrm{H}}
\newcommand{\grad}{\nabla}
\newcommand{\gradad}{\nabla_\mathrm{ad}}
\newcommand{\gradth}{\nabla_\mathrm{therm}}
\newcommand{\grade}{\nabla_\mathrm{e}}
\newcommand{\gradat}{\nabla_\mathrm{ad}'}
\newcommand{\Teff}{T_\mathrm{eff}}
\newcommand{\Ledd}{L_\mathrm{Edd}}
\newcommand{\kedd}{\kappa_\mathrm{Edd}}
\newcommand{\rphot}{R_\mathrm{phot}}
\newcommand{\intd}{\mathrm{d}}
\newcommand{\intdt}{\mathrm{d}t}
\newcommand{\Rcorr}{R_\mathrm{corr}}
\newcommand{\Lsurf}{L_\mathrm{surf}}
\newcommand{\hyphen}{\,--\,}

\def\bfnabla{{\mbox{\boldmath $\nabla$}}}

\renewcommand\bv{{\mbox{\boldmath $v$}}}

\newcommand\bn{{\mbox{\boldmath $n$}}}

\newcommand\br{{\mbox{\boldmath $r$}}}

\newcommand{\gtapprox}{\mathrel{\vcenter{
		\offinterlineskip\halign{\hfil$##$\cr
	>\cr\noalign{\kern2pt}\sim\cr\noalign{\kern-2pt}}}}}
\newcommand{\ltapprox}{\mathrel{\vcenter{
		\offinterlineskip\halign{\hfil$##$\cr
	<\cr\noalign{\kern2pt}\sim\cr\noalign{\kern-2pt}}}}}
	
\newcommand{\appropto}{\mathrel{\vcenter{
		\offinterlineskip\halign{\hfil$##$\cr
	\propto\cr\noalign{\kern2pt}\sim\cr\noalign{\kern-2pt}}}}}

\newlength{\apjcolwidth}
\newlength{\figwidth}
\newlength{\doublewide}
\renewcommand{\deleted}[1]{}

\begin{document}
\begin{CJK*}{UTF8}{gbsn}

\title{Numerical Simulations of Convective 3-Dimensional Red Supergiant Envelopes} 

\author[0000-0003-1012-3031]{Jared A. Goldberg}
\affiliation{Department of Physics, University of California, Santa Barbara, CA 93106, USA}

\author[0000-0002-2624-3399]{Yan-Fei Jiang(姜燕飞)}
\affiliation{Center for Computational Astrophysics, Flatiron Institute, New York, NY 10010, USA}

\author[0000-0001-8038-6836]{Lars Bildsten}
\affiliation{Department of Physics, University of California, Santa Barbara, CA 93106, USA}
\affiliation{Kavli Institute for Theoretical Physics, University of California, Santa Barbara, CA 93106, USA}

\correspondingauthor{J. A. Goldberg}
\email{goldberg@physics.ucsb.edu}

\begin{abstract}
 We explore the three-dimensional properties of convective, luminous
 ($L\approx10^{4.5}-10^{5}\Lsun$), Hydrogen-rich envelopes of Red Supergiants (RSGs) based on radiation hydrodynamic simulations in spherical geometry using $\Athena$. 
 These computations comprise $\approx30\%$ of the stellar volume, include gas 
 and radiation pressure, and
 self-consistently track the gravitational potential for the outer $\approx
 3\Msun$ of the simulated $M\approx15M_\odot$ stars. 
 This work reveals a radius, 
 $R_\mathrm{corr}$, around which the nature of the convection changes. 
 For $r>R_\mathrm{corr}$, though still optically thick, diffusion of 
 photons dominates the energy transport. Such a 
 regime is well-studied in less luminous 
 stars, but in RSGs, the near- (or above-) Eddington luminosity
 (due to opacity enhancements at ionization transitions) leads
 to the unusual outcome of denser regions moving outwards rather than
 inward. This region of the star also has a large amount of turbulent
 pressure, yielding a density structure much more extended than 1D
 stellar evolution predicts. This ``halo" of material will impact
 predictions for both shock breakout and early lightcurves of Type
 II-P supernovae. Inside of $R_\mathrm{corr}$, we find a nearly flat
 entropy profile as expected in the efficient regime of
 mixing-length-theory (MLT). Radiation pressure provides $\approx1/3$ of the support against gravity in this region. Our comparisons to MLT suggest a
 mixing length of $\alpha=3-4$, consistent with the sizes of
 convective plumes seen in the simulations. The temporal
 variability of these 3D models is mostly on the timescale of the
 convective plume lifetimes ($\approx300$ days), with amplitudes
 consistent with those observed photometrically. 
 \end{abstract}

\keywords{hydrodynamics --- radiative transfer --- convection --- stars: massive --- stars: supergiants}

\section{Introduction\label{sec:INTRODUCTION}}

As massive ($M\gtapprox10M_\odot$) stars leave the main sequence, they expand to become Red
Supergiants (RSGs), reaching radii of $\approx300-1200\Rsun$ and luminosities of 
$\approx10^4-10^{5.5}\Lsun$ \citep[e.g.][]{Levesque2006,Drout2012,Massey2021}, approaching the 
Eddington limit and receiving increasing hydrostatic support from radiation pressure.  
These stars are characterized by low-density convective hydrogen-rich envelopes with large scale heights ($H/r\approx0.3$) and sonic convection near their surfaces. They are intrinsically variable and pulsate in large-amplitude coherent modes (e.g. \citealt{Kiss2006,Soraisam2018,Chatys2019,Ren2019,DornWallenstein2020}) and their 3D nature is revealed to us in spectro-interferometric observations of nearby stars (e.g. \citealt{ArroyoTorres2015,Kravchenko2019,Kravchenko2021,Montarges2021,Norris2021}). 

It is a theoretical challenge to realistically model stars, or even parts of stars, in 3D. This is especially true when radiative transfer must also be 
simultaneously solved through a highly turbulent medium with large density variations over optical 
depths ranging from far above unity down to the radiating photosphere. This radiation hydrodynamic (RHD) challenge has been very 
well-addressed in cases where this region is close to plane-parallel, starting with the fundamental work for the Sun \citep{Stein1989,Stein1998}, and now ranging across the HR diagram 
\citep[e.g.][]{Trampedach2013,Trampedach2014a,Trampedach2014,Magic2013a,Magic2013b,Magic2015,Chiavassa2018,Sonoi2019}, building on earlier 2D RHD work (see \citealt{Ludwig1999} for an excellent summary). 
These 3D computations have yielded a physical understanding of the nature of RHD convection in this limit, and provide a quantitative ability to set the outer boundary condition in 1D stellar models \citep[e.g.][]{Trampedach2014a,Salaris2015,Magic2016,Mosumgaard2018,Spada2021} for $\log g\gtapprox1.5$, including for asteroseismic applications \citep{Mosumgaard2020}. 
While we have detailed understanding of the outer layers and quantitative surface relations for more compact, less luminous stars as guided by these works, such clarity has not been reached where the region requiring RHD calculations necessitates spherical geometry to capture large-scale plumes, and where the luminosity is locally super-Eddington.

In fainter giants, some of these aspects have been further addressed with global 3D simulations.
In Red Giant Branch (RGB) stars, simulations of the convective interior reveal relatively flat velocity profiles set by large-scale convective plumes, and large temperature and density fluctuations \citep{Brun2009}. These large-scale plumes extend up through the photosphere and produce granulation effects which can be interpreted by comparison of 3D models to interferometric data \citep[e.g.][]{Chiavassa2010b,Chiavassa2017}.
In Asymptotic Giant Branch (AGB) stars, 3D simulations have revealed additional insights about the 
pulsational and circumstellar structure, with nearly-spherical shock fronts from large-scale convective 
cells which also levitate material to radii at which they can form dust \citep[e.g.][]{Freytag2008, Freytag2017}. 
These simulations can then be used to, e.g., generate inner boundary conditions for 1D wind models 
\citep{Liljegren2018}, and interpret both interferometric and photometric observations \citep[e.g.][]{Chiavassa2018b,Chiavassa2020}.

In the luminous RSG regime, early simulations focused on surface turbulence and magnetic properties (e.g. 
\citealt{Freytag2002,Dorch2004}). Further simulations have been used to provide limb darkening 
coefficients and confirm the presence of large convective cells from interferometric observations of
Betelgeuse \citep{Chiavassa2009, Chiavassa2010}. \citet{Chiavassa2011a} provide photocentric noise 
models towards quantifying Gaia astrometric parallax uncertainties and explain the ``cosmic noise" 
impacting Hipparcos photometric measurements of Betelgeuse and Antares, while \citet{Chiavassa2011b} 
characterize microturbulence and macroturbulence parameters in grey- and frequency-dependent RSG 
atmosphere simulations. Further predictions from these models have been made with
radiation transfer post-processing with the software \texttt{OPTIM3D} \citep{Plez2013} 
and reveal the inability to define a single ``surface" responsible for setting the effective
temperature, $\Teff$. 

A unifying feature of theory and observations of RSGs is the turbulent, extended outer envelope
which manifests these inherently 3D convective properties. In 1D stellar evolution models, 
convection is conventionally handled by the Mixing Length Theory (MLT). 
The MLT approach derives from considering the fate 
of fluid elements as they move vertically a distance referred to as the mixing length
$\ell\equiv \alpha H$, where $\alpha$ is a free parameter which can be calibrated to observations or
by 3D simulations \citep{BohmVitense1958,Henyey1965,Cox1968}. 
Especially in Red Giants and Supergiants, mixing length assumptions, especially the value of $\alpha$ (and
assumptions relevant to the structure and location of convective boundaries, which we will not explore in 
this work) strongly influence the stellar radii and $\Teff$ \citep[e.g.][]{Stothers1995,Meynet1997,Massey2003,Meynet2015}. 
While empirical constraints are useful, even crucial, for producing RSG models which match 
observed stars in luminosity, $L$, and $\Teff$ \citep[e.g.][]{Chun2018}, a first-principles calibration of MLT to 3D 
simulations of RSG envelopes remains an open channel for theoretical progress in characterizing
the nature of convection in these very luminous objects.

The turbulent RSG envelope also plays a crucial role at the end of the star's life, as a strong 
shock emerges from the collapsed core and propagates rapidly through the envelope. Such
explosions result in Type II-P SNe with $\simeq\!100$-day plateaus in their lightcurves whose properties
depend on the envelope structure, and especially the progenitor radius, ejected mass, explosion energy and
$^{56}$Ni mass (e.g. \citealt{Popov1993,Kasen2009,Sukhbold2016}). The exact initial mass range of 
stars exploding as Type II-Ps is still a matter of significant debate \citep[the so-called ``RSG problem'', e.g.][]{Smartt2009,Smartt2015,Davies2018,Kochanek2020,Davies2020a,Davies2020b}. 
If the RSG radius is known at the moment of explosion, then light curve modeling can be used to 
constrain the ejected mass \citep{Goldberg2019,Martinez2019, Goldberg2020}, with some sensitivity to the 
pulsation mode and phase at the time of explosion (see discussion in \citealt{Goldberg2020a}). 
However, if the progenitor radius is unknown, very different stellar properties can yield identical 
lightcurves and photospheric velocities after the first $\approx30$ days 
\citep{Dessart2019,Goldberg2019}, limiting our ability to infer masses and explosion energies solely from 
these observations. 

Early Supernova observations can assist with breaking these degeneracies, but doing so is hampered 
by our lack of understanding of the density structure of the outermost RSG layers responsible for 
the early time emission \citep[see, e.g.][]{Morozova16}. In addition, Type II-P SNe frequently exhibit 
luminosities in excess of explosion models that assume a simple stellar photosphere (e.g. 
\citealt{Morozova2017,Morozova2018}). This early excess is often attributed to interaction between the 
SN ejecta and the progenitor's outgoing wind \citep[e.g.][]{Moriya2018} or ejecta from pre-SN outbursts
\citep{Fuller2017,Morozova2020}, and poses challenge in cleanly interpreting these early phases of SN 
evolution \citep[see, e.g.][]{Hosseinzadeh2018}. It is also possible that these discrepancies are 
because the density structure in the vicinity of the photosphere is simply not well-described by 
conventional 1D stellar models. One important long-term goal of our effort is to better constrain the 
role of the 3D gas distribution in early SN emission. 

This paper is organized as follows: In \S\ref{sec:background}, we describe motivating
expectations for the 3D regime we aim to explore, making use of Modules for Experiments in Stellar 
Astrophysics \citep[\MESA][]{Paxton2011,Paxton2013,Paxton2015,Paxton2018,Paxton2019} to illustrate the 
importance of a proper 3D treatment of RSG envelopes. In \S\ref{sec:athenasetup} we describe our 3D 
\Athena\ \citep{Stone2020} RHD setup for RSG envelopes, and in \S\ref{sec:3Dprops} we discuss the 
convective properties of these envelopes, comparing where possible to findings of earlier 3D 
RSG models. We then compare our 3D envelope models to predictions from MLT where appropriate 
(\S\ref{sec:3Dto1D}). Finally, we discuss our results and comment on future directions in \S\ref{sec:conclusions}.

\section{Properties of 1D Red Supergiant Models and Open Challenges\label{sec:background}}
\label{sec:mesa}
\begin{figure}
\includegraphics[width=\columnwidth]{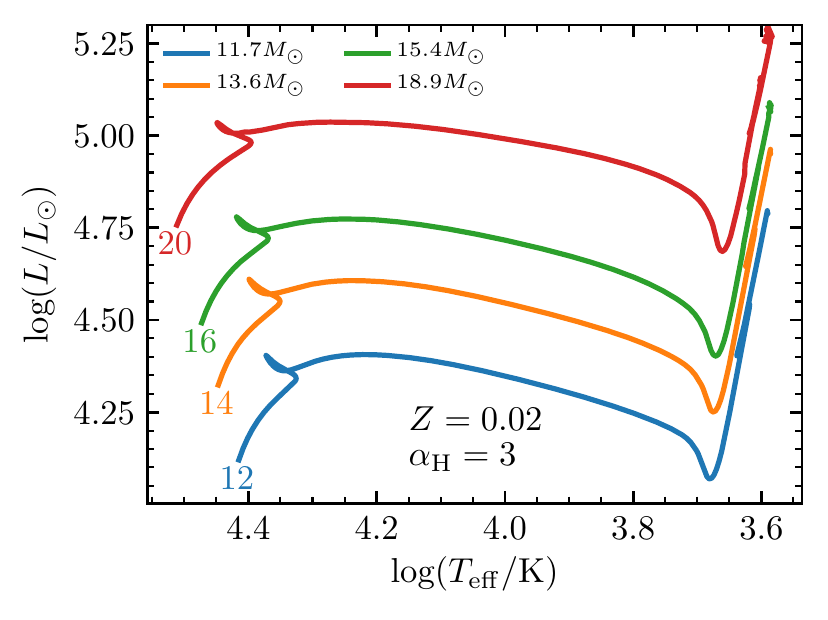} 
\caption{H-R diagrams of non-rotating \MESA\ models with initial masses of $M_i/M_\odot$=12 (blue), 14 (orange), 16 (green), and 20 (red), from the main sequence through core Si burning. Final masses are given in the legend.}
\label{fig:HR0}
\end{figure}

For our initial exploration, we constructed a suite of solar-metallicity ($Z=0.02$) models in \MESA, 
following the test suite case \texttt{make\_pre\_ccsn\_IIp} in revision 15140, shown 
in Fig.~\ref{fig:HR0} from the onset of core H burning through the end of core Si burning. Our fiducial non-rotating models have modest exponential overshoot with overshooting parameter $f_\mathrm{ov}=0.016$,
a wind efficiency of $\eta_\mathrm{wind}=0.2$ using the 
\textquotesingle\texttt{Dutch}\textquotesingle\ scheme in \MESA\ 
\citep{Nugis2000,Vink2001,Glebbeek2009}, core mixing length $\alpha_\mathrm{c}=1.5$ in 
regions where the H fraction $X_\mathrm{H}\leq0.5$, and mixing length $\alpha_\mathrm{H}=3$ in 
the H-rich envelope ($X_\mathrm{H}>0.5$). 
These parameters were chosen to be similar to those of the Type IIP Supernova progenitor models in \citet{Paxton2018}, 
motivated also by the findings of \citet{Farmer2016}.

\begin{figure*}
\centering
\includegraphics[width=0.48\textwidth]{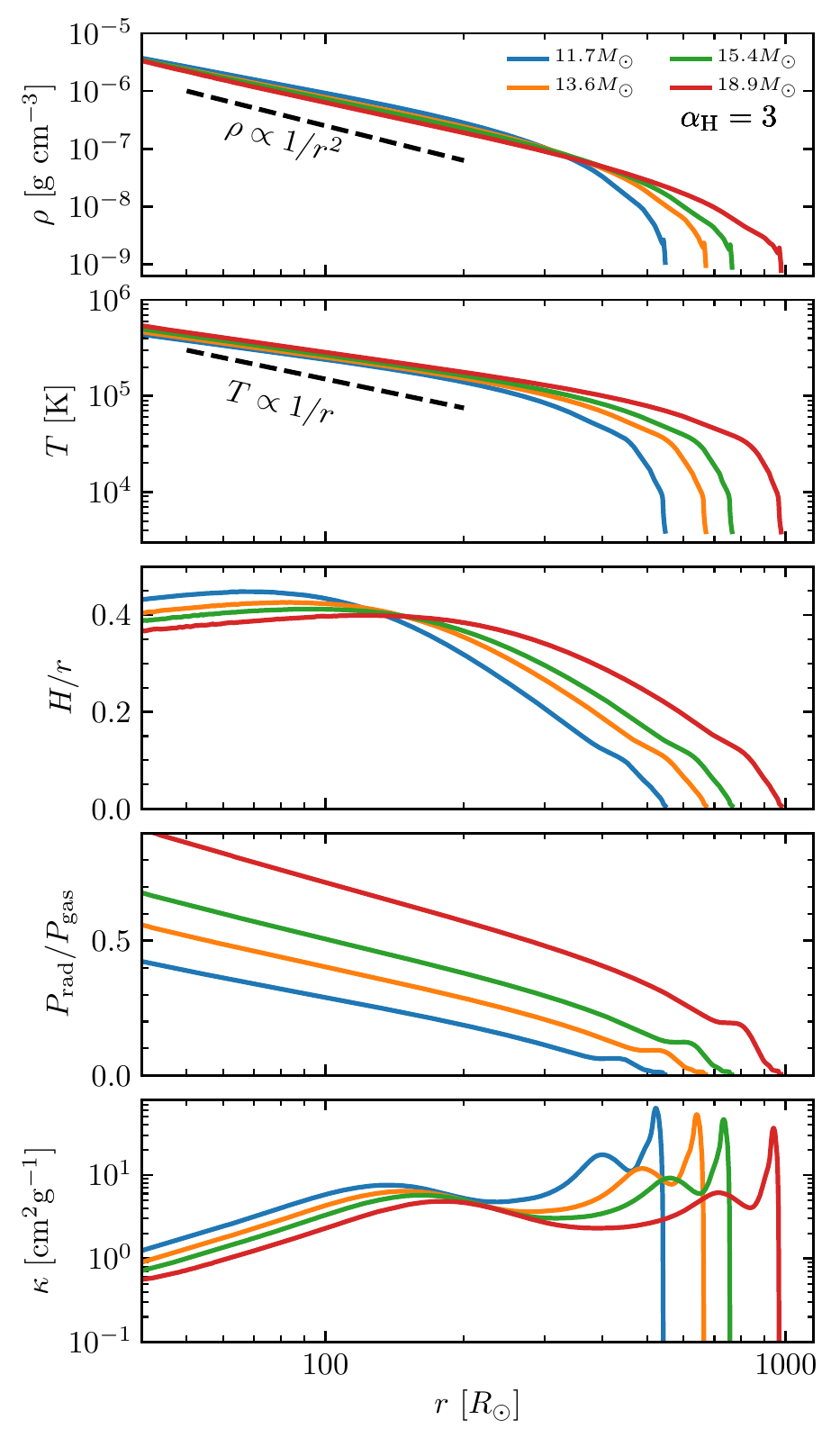} 
\includegraphics[width=0.48\textwidth]{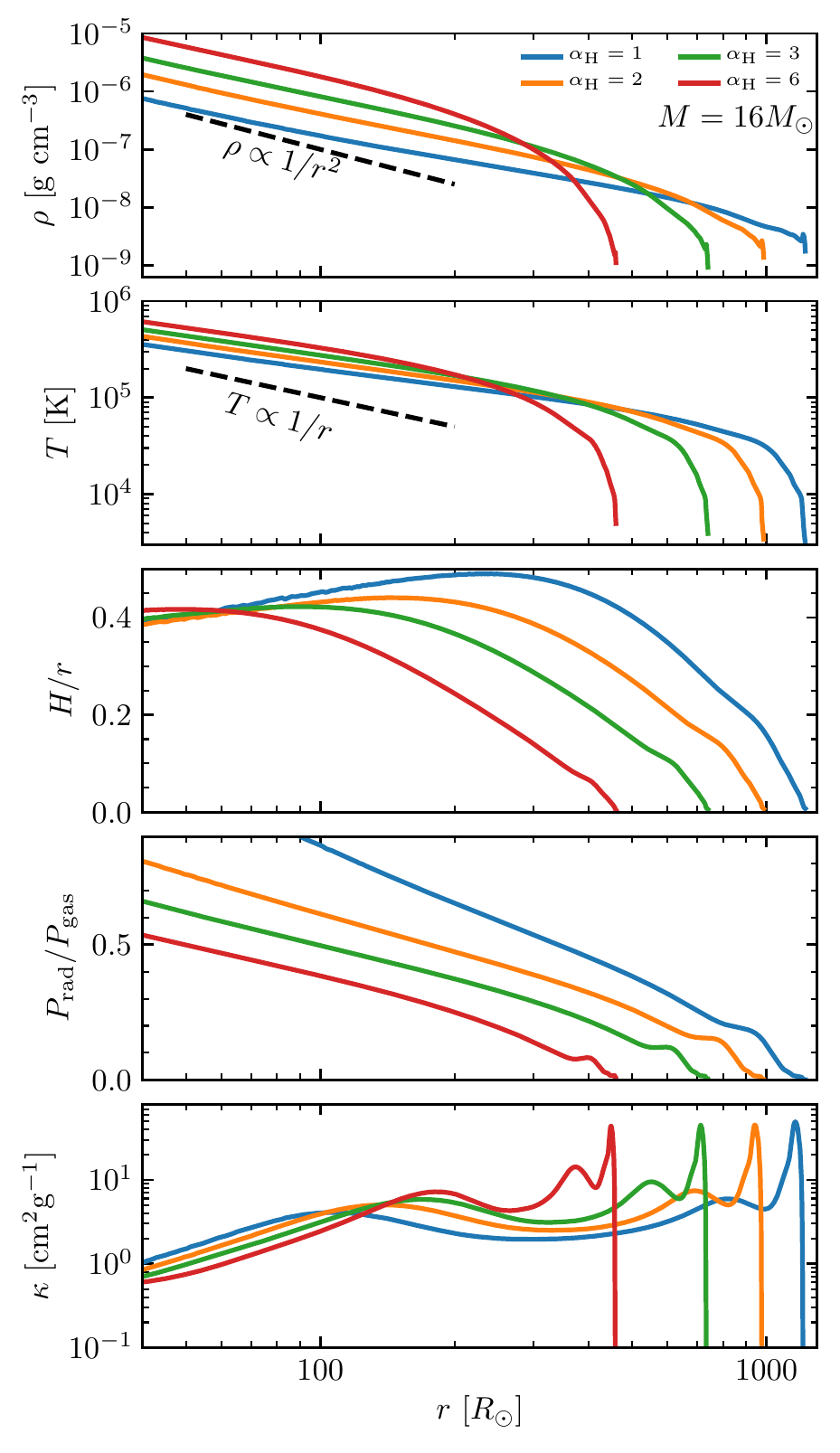} 
\caption{Top to bottom: Density, temperature, pressure scale height divided by radius, radiation to gas pressure ratio, and opacity as a function of radial coordinate $r$ in $Z=0.02$ RSG models. Left: Initial masses of $M_i/M_\odot$=12 (blue), 14 (orange), 16 (green), and 20 (red), all with $\alpha_\mathrm{H}=3$. Right: Masses of $M=16M_\odot$, varying the mixing length in the Hydrogen-rich envelope, $\ah$, for $\alpha_\mathrm{H}=1$ (blue), $\alpha_\mathrm{H}=2$ (orange), $\alpha_\mathrm{H}=3$ (green), and $\alpha_\mathrm{H}=6$ (red); here winds were neglected to isolate the effects of varying $\ah$, leading to the slight differences between the green lines in the left and right panels.}
\label{fig:panels1}
\end{figure*}

The left panels of Figure~\ref{fig:panels1} show the structure of four model RSG envelopes at the 
end of core C burning (central $X_\mathrm{C}<10^{-6}$) with initial masses ranging from 12 to 20 
$M_\odot$. The x-axis excludes the He core, which is always inside of $r=10R_\odot$ for all models. 
Through the envelope, the density falls by 3-4 orders of 
magnitude, nearly matching $\rho\propto1/r^2$ through most of the inner envelope. The pressure 
scale height, $H=P/\rho g$, is large due to the weak gravity in the envelope, with $H/r\approx0.3$ even 
at the half-radius coordinate. The envelope is fully convective, and both radiation 
pressure and gas pressure contribute significantly to the total pressure, with gas pressure 
dominating near the surface. Additionally, the opacity is very large throughout the envelope, 
dominated by opacity peaks from H and He ionization transitions inside the convective region. 

Where convection is ``efficient", $\grad$ is nearly $\grade\approx\gradad$ and the fluid structure follows the adiabat.
There are two senses in which convection is said to be inefficient.
When the convection is inefficient in 
the superadiabatic sense (i.e. $\grad\gg\grade$, where $\grad=d\ln{T}/d\ln{P}$ and $\grade$ is the internal 
$\grad$ of a convective parcel; see Table~\ref{tab:grads} in Appendix~\ref{sec:AppendixB}), a rising fluid element will be hotter than the surrounding medium, and it 
will accelerate as it moves outwards in order to carry the flux. The stellar entropy profile thus declines. 
The convection can also be inefficient in the radiative sense, or ``lossy", when a convective fluid parcel has sufficient time to
radiate its internal energy to the cooler surrounding as it rises. 
In a medium with $\Pgas\gg\Prad$, the optical depth at which radiation is able to contribute 
significantly to the energy transport and lossy convection is expected is $\tau<\taut$, where  
 \begin{equation}
 \taut\approx\frac{\Prad}{\Pgas}\frac{c}{\vc},
 \label{eq:taut}
 \end{equation}
where $c$ is the speed of
light, and $\vc$ is the radial component of the convective velocity. The factor of 
$\Prad/\Pgas$ comes from the fact that in the gas-pressure-dominated region near the cool 
stellar surface the parcel must evacuate the radiation field $\Prad/\Pgas$ times in order to 
carry the same flux by radiation as convection \citep{Kippenhahn}. For $\tau>\taut$ where a parcel is 
unable to lose heat to radiation, $\grade\approx\gradad$. We will note later, in \S\ref{sec:fund3D}, the close relationship between $\tau/\taut$ and the more commonly-discussed convective efficiency parameter, $\gamma$.

 In MLT, $\grad-\grade$ is directly related to the mixing length $\ell=\alpha H$ by \citep{Kippenhahn}
 \begin{equation}
     F_\mathrm{conv}=\rho c_{P}T \sqrt{gQ}\frac{\ell^2}{\sqrt{\nu}}H^{-3/2}(\nabla-\nabla_\mathrm{e})^{3/2},
     \label{eq:mltflux}
 \end{equation}
 where $F_\mathrm{conv}$ is the flux carried by convection, $Q=-\mathrm{D}\ln T/\mathrm{D}\ln\rho=(4-3\beta_P)/\beta_P$ where $\beta_P=\Pgas/(\Prad+\Pgas)$ for
 an equation of state (EOS) made up of radiation and gas, $\nu=8$ following 
 \citet{Henyey1965} and others, and $c_P$ is the specific heat at constant pressure. 
 
So as to explore the dependence of the RSG envelope structure on the mixing length $\ah$, we constructed 
additional 16$M_\odot$ RSG models varying $\alpha_{H}$ from 1 to 6. 
In these, we neglect mass loss due to winds ($\eta_\mathrm{wind}=0.0$)
and vary $\alpha_\mathrm{H}$ away from the fiducial value of $\ah=3$ 
only at the end of core He burning in order to ensure that the 
resulting models have comparable core masses, $M_\mathrm{c,He}=5.2\Msun$, and luminosities, $\log(L/\Lsun)=5.06$. The structure of these models is 
shown in the right panels of Figure~\ref{fig:panels1}. Lower values of $\ah$ 
produce models with larger radii, lower densities, and lower temperatures throughout 
the envelope.

\begin{figure*}
\centering
\includegraphics[width=0.48\textwidth]{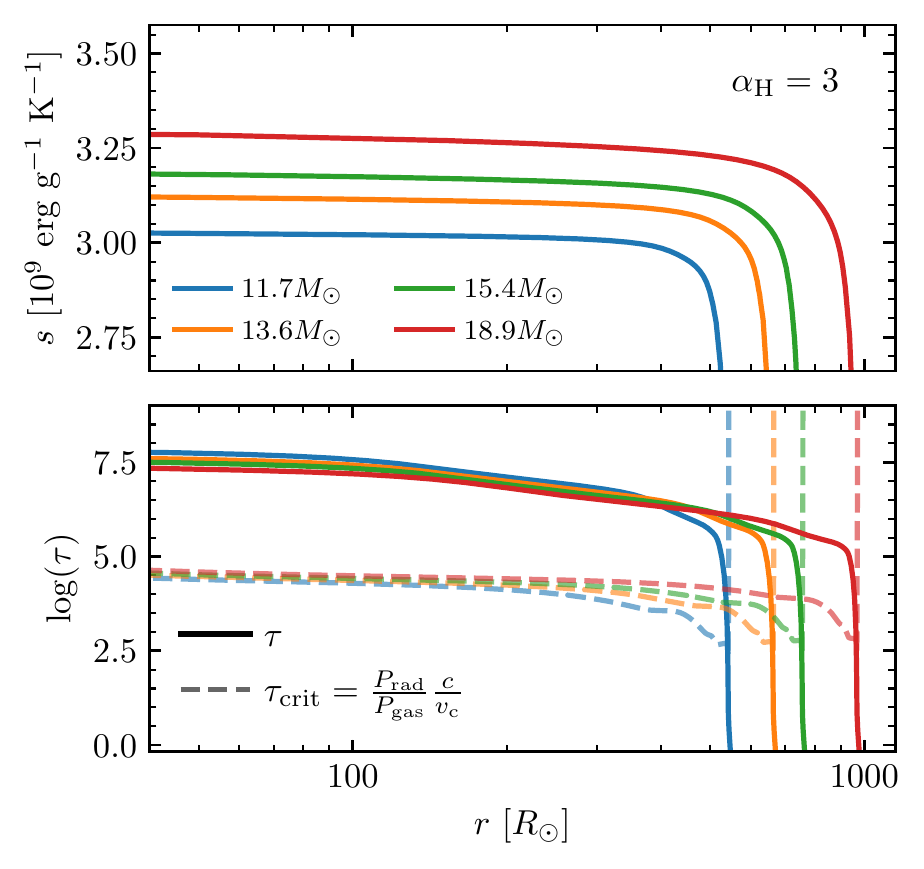} %
\includegraphics[width=0.48\textwidth]{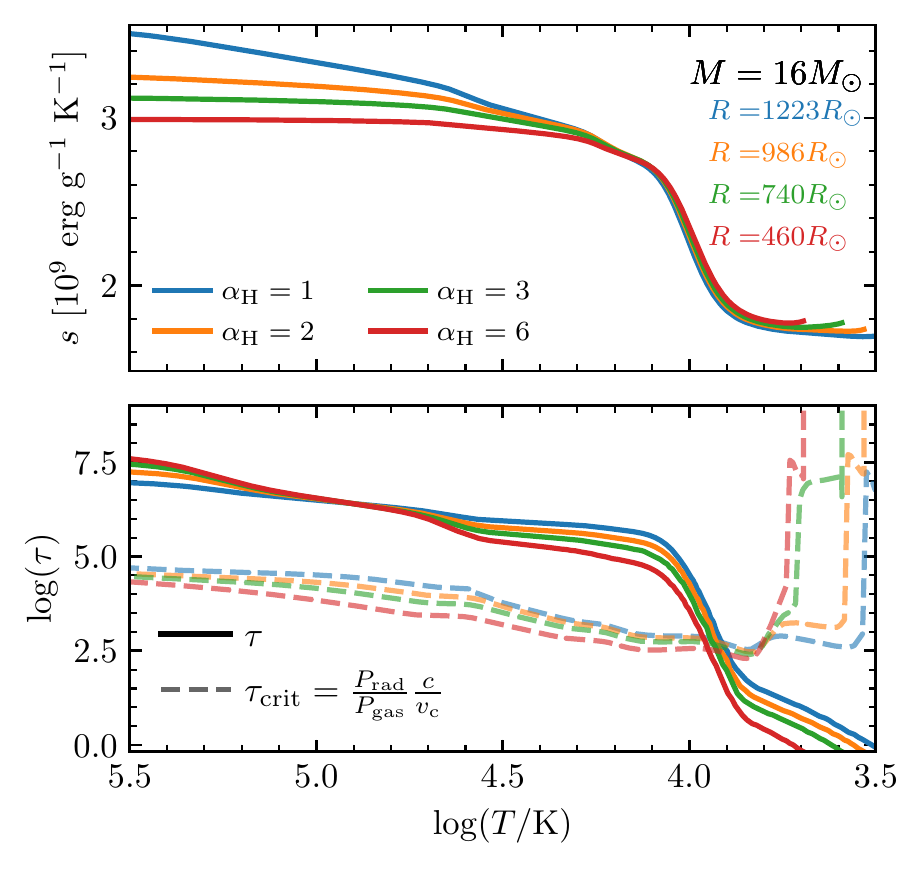} 
\caption{Specific entropy (upper panels) and optical depths (lower panels) for RSG models of different initial mass and $\ah=3$ (left panels), and varying $\ah$ with $M=16\Msun$ (right panels). The optical depth remains much higher than $\taut$ (dashed lines) until $\tau\approx300$ near $T\approx10^4$\,K.}
\label{fig:entropy2}
\end{figure*}

The upper panels of Fig.~\ref{fig:entropy2} show the specific entropy, $s$, 
profiles for models varying the initial mass (left) and $\ah$ (right) at the end of core C burning. 
The lower panels compare $\taut$ (dashed lines) to 
the optical depth $\tau$ (solid lines). The transition to lossy convection with radiation-dominated 
transport typically occurs around $T\approx10^4$K and $\tau=\taut\approx300$, which is near the H 
opacity peak seen in Fig.~\ref{fig:panels1}. At that location, the peak in opacity and 
large luminosity implies $L\gg\Ledd$ there, a critical distinction for RSG models compared to 
main-sequence, RGB, or AGB stars. 
As $T$ approaches $\Teff$, $\vc$ declines to zero in a very thin radiative region above the convection 
zone. The green models are comparable between the left and right panels, with the only substantive difference being the inclusion of mass loss in the left panel leading to a slightly lower core mass, $M_\mathrm{c,He}=4.9\Msun$, and luminosity, $\log(L/\Lsun)=5.02$. 

Varying initial mass increases the luminosity and thereby $s$, with relatively flat entropy profiles that 
begin to decline near the surface.  Decreasing $\ah$ decreases the efficiency of the convection, causing a 
steeper temperature gradient and an entropy decline. Larger mixing lengths correspond to more 
efficient convection and produce higher $\Teff$. For a given luminosity, this leads to different radii with
varying $\ah$, from $R=460\Rsun$ when $\ah=6$ to $R=1223\Rsun$ when $\ah=1$, despite 
comparable envelope masses and luminosities. 

The assumed mixing length thus plays a dual role in determining the stellar 
structure. Foremost, the entropy profile declines even where $\tau\gg\taut$, especially for lower $\ah$, 
suggesting true superadiabatic convection with nonnegligible $\grad-\grade$.  
The choice of $\ah$ influences $\grad-\grade$ via Eq. \eqref{eq:mltflux} for a given 
$F_\mathrm{conv}$, and therefore determines the deviation of the temperature profile from the adiabat. 
Secondly, $\ah$ determines the adiabat on which the envelope sits. This effect can also be seen in models of cool stars more generally \citep[e.g.][]{Stothers1995,Meynet1997,Massey2003,Meynet2015} and is pronounced where convection occurs over orders of magnitude in radius, such as in cool giants.
Running a further suite of models where we varied the location where the mixing length coefficient changes from a fixed $\alpha_\mathrm{c}$=1.5 to variable $\ah$ at different temperature coordinates, rather than setting the transition to be
at the H-He interface as in our fiducial models, 
we find that changing $\ah$ in the lossy outer envelope below a 
few times $10^4$K (where $\tau\ltapprox\taut$) is what primarily determines the outer radius of the star, 
as the entropy decline in that region is fixed (as seen in the upper right panel of 
Fig.~\ref{fig:entropy2}). 
Since $\rho\appropto1/r^2$, the stellar radius determines the density at the 
base of the envelope, the radiation to gas pressure ratio, the entropy, and thereby the adiabat. So 
even though less efficient convection at lower $\ah$ would predict a steeper radial temperature profile
for fixed inner boundary, this is more than offset by the fact that the entropy deep within the 
envelope is larger for lower $\ah$. 
Although $\ah$ has been constrained for stellar models where 
$\Prad\ll\Pgas$ and $H/r\ll1$ throughout their convective regions \citep[see, e.g.][]{Trampedach2014,Magic2015,Sonoi2019}, the `true' value of $\ah$ in the RSG 
envelope regime has never been calibrated to 3D simulations. Comparisons of 1D stellar models to observed 
RSG populations suggests $\ah\approx2-3$ in different environments based on their
location on the HR diagram, and in particular their effective temperatures (e.g. 
\citealt{Ekstrom2012,Georgy2013,Chun2018}).\footnote{See also the discussion by \citet{Joyce2020} of 
how MLT uncertainties bear on stellar evolutionary and hydrodynamical models of $\alpha$ Ori compared 
to asteroseismic observations.}  

 In MLT, the convective velocity $\vc$ is related to the 
 superadiabaticity and the mixing length by
 \begin{equation}
     \vc^2=gQ\left(\grad-\grade\right)\frac{\ell^2}{\nu H}.
     \label{eq:mltv}
 \end{equation}
 Where $\tau>\taut$, a fluid parcel retains most of its heat and
 $\grade\approx\gradad$.
 Note that superadiabatic convection with $\grad>\grade\approx\gradad$ leads to an increase in the convective velocity, while lossy convection yields a decrease in the convective velocity required to carry the flux as $\grade$ deviates from $\gradad$ and approaches $\grad$.

\begin{figure}
\centering
\includegraphics[width=\columnwidth]{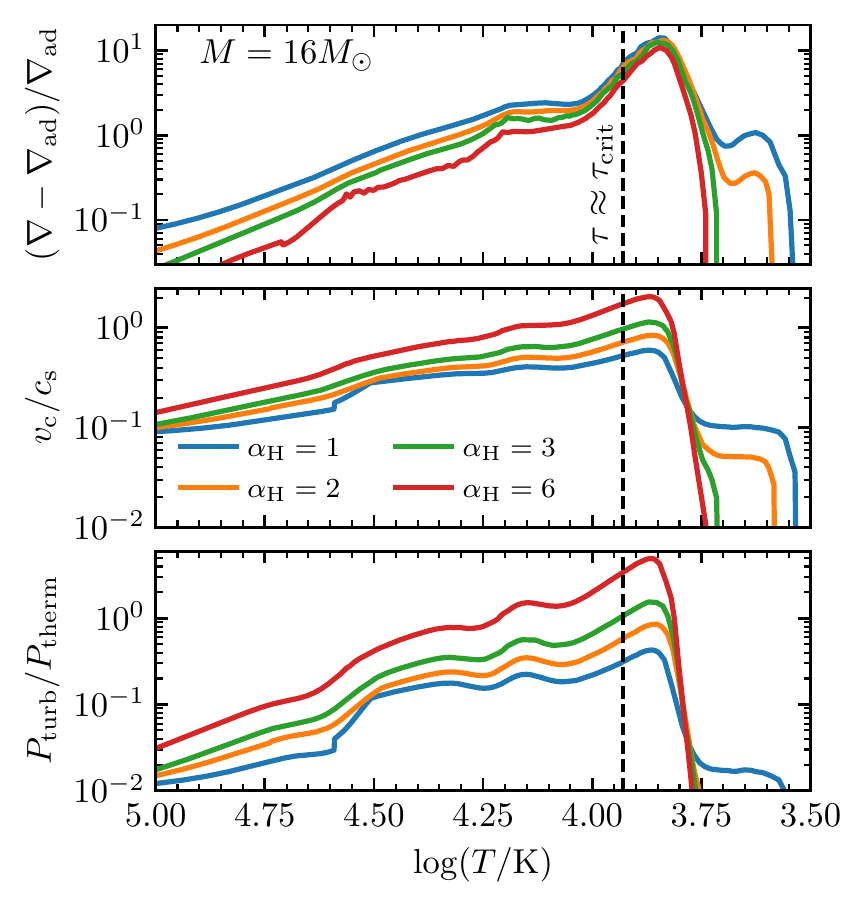} 
\caption{The superadiabaticity (upper panel), convective Mach number (middle panel), and estimated turbulent pressure (lower panel) versus log($T/$K) for $M=16\Msun$, $Z=0.02$ RSG models as $\ah$ varies. The vertical dashed lines indicate where $\tau\approx\taut$ in these models.}
\label{fig:superadiabaticity3}
\end{figure}

 Fig.~\ref{fig:superadiabaticity3} shows the superadiabaticity $(\grad-\gradad)/\gradad$ (upper panel) and convective Mach number (middle panel) as a function of temperature coordinate for four $16\Msun$ models with varying $\ah$. 
 As the superadiabaticity becomes large, particularly for larger $\ah$, convective velocities become nearly supersonic. 

In the plane parallel limit, the turbulent pressure term needed to incorporate the effects of the
3D Reynolds stress in a radial 1D model is $\Pturb=\rho\vr\vr\approx\beta\rho\vc^2$ 
up to a prefactor $\beta$ typically assumed to be unity \citep{Henyey1965}.\footnote{
In a plane parallel atmosphere where the z-direction is identified with radial gravity, the radial component of the gradient of $\rho\bv\bv$ is equal to the gradient 
of $\rho\vr\vr$ when deriving $\Pturb$ from the Euler equations. 
However, in spherical polar geometry the gradient of  $\rho\bv\bv$ yields geometric terms 
$(\rho\vtheta\vtheta+\rho\vphi\vphi)/r$ which contribute to the momentum equation \citet{LandauLifschitz}. 
These terms are a small correction when $H\ll r$, which is not strictly the case in the RSG regime, 
or could vanish if $2\vr\vr-(\vtheta\vtheta+\vphi\vphi)\approx0$.} 
This quantity is shown 
in the lower panel of Fig.~\ref{fig:superadiabaticity3} for $\vc$ given by \MESA\ assuming $\beta=1$. Moving towards the stellar surface, the expected
turbulent pressure rises, even exceeding the thermal pressure ($\Ptherm=\Pgas+\Prad$) in the cooler ($T\ltapprox10^4$ K) regions of the $\ah\geq3$ 
models. Due to the intrinsically 3D nature of large-scale convection and the resulting turbulent pressure, 
the handling of this large expected pressure contribution is another way in which 3D results can guide 1D models. 

\begin{figure}
\centering
\includegraphics[width=\columnwidth]{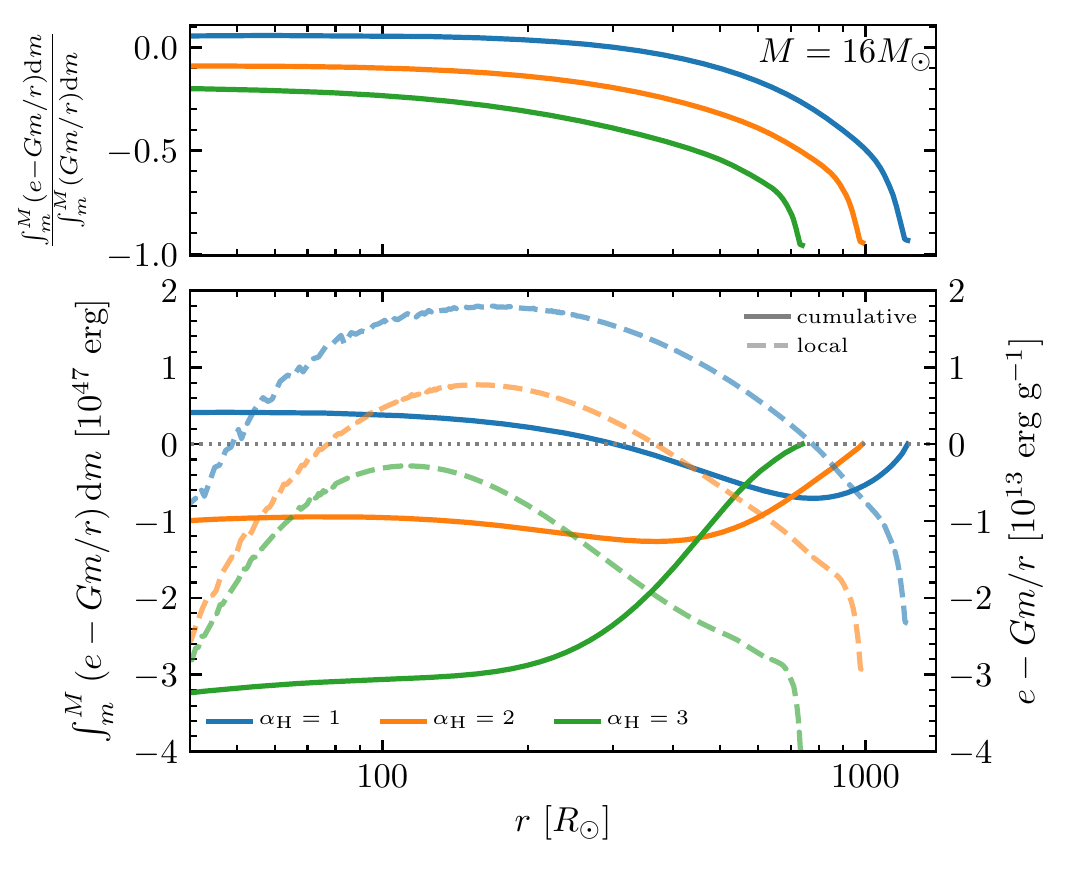} 
\caption{Upper panel: The ratio of the cumulative internal energy to the magnitude of the cumulative potential energy, integrated from the surface. Lower panel: Specific (solid) and cumulative (dashed) total energy (IE + PE) calculated from the surface inward in the envelope of 16 $M_\odot$ RSG models with $\alpha_\mathrm{H}$=1,~2,~3.}
\label{fig:energyint5}
\end{figure}

Moreover, the envelopes of these models are only very loosely gravitationally bound. The lower panel of 
Fig.~\ref{fig:energyint5} shows the local total energy (dashed lines) and cumulative total energy 
integrated from the surface inwards (solid lines). The upper panel shows the ratio of the cumulative total 
energy to the gravitational energy. The kinetic energy assuming $v=\vc$ is neglected, as it only 
contributes only 
a few times $10^{45}$ erg in total for these models. As seen in 
the upper panel, the gravitational energy and the internal energy nearly cancel, and for our $\ah=1$ 
model the internal energy exceeds the gravitational binding energy inside the envelope. This demonstrates the precariousness of these RSG envelopes, and why they can become unbound even from small energy deposited there from direct collapse of the 
He core to a black hole (e.g. \citealt{Nadezhin1980}; \citealt{Coughlin2018}). This also highlights the importance of incorporating the envelope's self-gravity in our 3D calculations.

\section{3D Model Setup and Equilibration}\label{sec:athenasetup}

\begin{table*}\label{tab:models}
\begin{center}
\begin{tabular}{| c | c | c | c | c | c | c | c | }
\hline Name & $R_\mathrm{IB}/\Rsun$ & $R_\mathrm{out}/\Rsun$ & heat source & resolution ($r\times\theta\times\phi$) & duration & $m_\mathrm{c}/\Msun$ & $M_\mathrm{final}/\Msun$ \\ \hline
RSG1L4.5 & 400 & 22400 & ``hot plate" & $384\times128\times256$ & 5865 d & 12.8 & $16.4$ \\ \hline

RSG2L4.9 & 300 & 6700 & fixed $L$ & $256\times128\times256$ & 5766 d & $10.79$ & $12.9$ \\ \hline

\end{tabular}
\end{center}
\caption{Simulation properties, including inner boundary ($R_\mathrm{IB}$), outer boundary ($R_\mathrm{IB}$), heat source (as described in the text), resolution, run duration, core mass $m_\mathrm{c}$, and total mass at the simulation end ($M_\mathrm{final}$). All models have $\theta=\pi/4-3\pi/4$ and $\phi=0-\pi$, with $\delta r/r\approx 0.01$, and we restrict our analysis to material outside 450$\Rsun$. The naming scheme indicates $\log(L/\Lsun)$.}
\end{table*}

\subsection{Model Setup in \Athena}
To explore the 3D convective properties of RSGs, we constructed two large-scale simulations using \Athena. 
For these simulations, we use spherical polar coordinates with 128 uniform bins in polar angle 
$\theta$ from $\pi/4-3\pi/4$ and 256 bins in azimuth $\phi$ from $0-\pi$ with periodic boundary 
conditions in $\theta$ and $\phi$, covering 70.6\% of the face-on hemisphere (i.e. solid angle $\Omega=1.41\pi$).
Outside of the simulation domain, \Athena\ uses ghost zones to enforce its boundary conditions (see \citealt{Stone2020} for more details). For the ``periodic" boundary in $\theta$, the ghost zones from $\pi/4$ ($3\pi/4$) are copied from last active zones around the $3\pi/4$ ($\pi/4$) boundary, so that the mass and energy flux across the theta boundary is conserved. Although the spherical polar grid in \Athena\ can in principle include the whole sphere, such a setup will cause a timestep that is too small to perform these simulations. That is why we only cover the polar region between $\pi/4$ and $3\pi/4$, which is designed to represent a large typical wedge of the star. There 
are 2 options for a boundary condition in order to conserve mass and energy in the $\theta/\phi$ direction.
The method described here is preferred over a reflective
boundary condition, which will lead to ``splashback" (as is seen at the inner boundary).
\Athena\ solves the ideal hydrodynamic equations coupled with the time-dependent, 
frequency-integrated radiation transport equation for specific intensities over discrete angles
\citep{Jiang2014,Jiang2021}. We adopt the spherical polar angular system as defined in Section 3.2.4 of \cite{Jiang2021} with 120 total angles per grid for the specific intensities. In this 
initial work, we consider a non-rotating stellar model and neglect magnetic fields. This is likely a safe assumption, as the envelope rotation reduces dramatically as the stars ascend the Hayashi track after core H depletion, though some RSG envelopes may have non-negligible rotation due to interaction or a merger with a companion (see, e.g., \citealt{Joyce2020}).

The RHD equations are \citep{Jiang2021}:

\vspace{-.2in}
\begin{equation} \label{eq:RHD}
\begin{split}
\frac{\partial\rho}{\partial t}+\bfnabla\cdot(\rho\bv)&=0,\\
\frac{\partial(\rho\bv)}{\partial t}+\bfnabla\cdot({\rho\bv\bv+{{\sf\Pgas}}}) &=-\mathbf{G}_r-\rho\bfnabla\Phi, \\
\frac{\partial{E}}{\partial t}+\bfnabla\cdot\left[(E+\Pgas)\bv\right]&=-c G^0_r -\rho\bv\cdot\bfnabla\Phi,\\
\frac{\partial I}{\partial t}+c\bn\cdot\bfnabla I&=S(I,\bn),
\end{split}
\end{equation}
where $\rho$ is the gas density and $\bv$ is the flow velocity. The gas pressure tensor and scalar are given by ${\sf\Pgas}$ and $\Pgas$, respectively. The total gas energy density is $E=E_g+\rho v^2/2$, 
where $E_g=3\Pgas/2$ is the gas internal energy density. 
Source terms $G^0_r$ and $\mathbf{G}_r$ are the time-like and space-like components of the radiation four-force
\cite{Mihalas1984}. The frequency-integrated intensity $I$ is a function of time, spatial coordinate, and photon propagation direction $\bn$. 

The mass in the simulation domain is not negligible, and because the envelope is expected to be only 
loosely bound, it is important to include an accurate gravitational acceleration, which we take to be 
spherically symmetric, with $-\nabla\Phi=-Gm(r)/r^2$. Here $G$ is the gravitational constant, $r$ is the radial 
coordinate, and $m(r)$ is the total mass inside $r$. We calculate $m(r)$ as the sum of the ``core'' 
mass interior to the inner boundary (IB) and the total mass between the IB and $r$ at each time step.\footnote{
An exploratory simulation did not include the self-gravity of the material within our simulation domain, instead using only a fixed mass from inside the inner boundary. In that simulation, the envelope rapidly expanded to $\rphot$$>$$3000\Rsun$ with a sharp increase in mass in the simulation domain coming from the IB, and never reached a quasi-hydrostatic convective steady state.}
The gas temperature is $T=(\Pgas\mu m_\mathrm{p})/(k_\mathrm{B}\rho)$, where $k_B$ is the Boltzmann constant,
and $m_p$ is the proton mass, with mean molecular weight $\mu=0.643$ to match our \MESA\ models. A 
radiation temperature $T_r$ can be calculated from the radiation energy density $E_r$ included in the 
$G^0_r$ source term as $T_r=(E_r/a_r)^{1/4}$ where $a_r=4\sigma_\mathrm{SB}/c$ is the radiation constant
and $\sigma_\mathrm{SB}$ is the Stefan-Boltzmann constant; this is typically, but not necessarily, identical to $T$.

To calculate the radiation energy and momentum source terms, the lab frame intensity $I(\bn)$ with angle $\bn$
is first transformed to the co-moving frame intensity $I_0(\bn_0)$ with angle $\bn_0$ via Lorentz transformation \citep{Mihalas1984,Jiang2021}.
The source terms describing the interactions between gas and radiation in the
comoving frame are 
\begin{equation}
S_0(I_0,\bn_0)=c\rho\kappa_{aP}\left(\frac{ca_rT^4}{4\pi}-J_0\right)+c\rho(\kappa_s+\kappa_{aR})\left(J_0-I_0\right),
\label{eqn:radsource}
\end{equation}
where $\kappa_{aR}$ and $\kappa_{aP}$ are the Rosseland and Planck mean absorption opacities determined by interpolation of the OPAL opacity tables \citep{Iglesias1996},
and $\kappa_s$ is the electron scattering opacity, all evaluated in the comoving 
frame. The angular quadrature of the intensity in the 
co-moving frame is $J_0=\int I_0(\bn_0)d\Omega_0/(4\pi)$. After the specific intensities $I_0(\bn_0)$ are updated in the co-moving frame, they are Lorentz transformed back to the lab frame. The radiation momentum and energy source terms $G^0_r$ and $\mathbf{G}_r$ are calculated by the differences between the angular quadratures of $I(\bn)$ in the lab frame before and after adding the source terms. See \cite{Jiang2021} for more details of the implementation. The hydrodynamic equations are solved using the standard Godunov method in \Athena\ \citep{Stone2020}. Similar numerical methods and setup have been successfully used to model stellar envelopes in different locations of the HR diagram \citep{Jiang2015, Jiang2018}.

\subsection{RSG Setup and Model Evolution}
We used the NASA supercomputer Pleiades to run two 3D RHD simulations. 
Each run takes about two months to finish with 80 skylake nodes in Pleiades.
For this study, we motivate our initial and boundary conditions with the fiducial $15.4\Msun$, $Z=0.02$, $\ah=3$ model at the end of core C burning discussed in \S\ref{sec:mesa} (shown in green in the left panels of Fig.~\ref{fig:panels1}). 
Our first model, referred to hereafter RSG1L4.5, is initialized in 3D by assuming a purely 
radiative envelope with luminosity equal to the radiative luminosity at $r=400\Rsun$ in the \MESA\ model (which is a few \% of the total luminosity). The mass and radius 
of the IB are $400\Rsun$ and 12.8$\Msun$. To generate the initial conditions, the temperature ($T=7.19\times 10^4$ K) at
the IB is first set to equal the $400\Rsun$ 
coordinate in the \MESA\ model, and density ($\rho=5.45\times 10^{-8}$ g/ cm$^{-3}$) selected to approximately recover the $15.4\Msun$ 
total mass. To perturb from the radiatively stable initial conditions and supply the convective luminosity, we increase the temperature at the IB by $10\%$ compared with the initial condition (a ``hot plate"), while density is fixed and velocity is reflective at the inner boundary, akin to the setup of \citet{Jiang2018}. This boundary condition produces a radiative layer near the bottom with the desired luminosity, which causes the envelope away from the bottom boundary to be convective. In this setup, we do not know in advance what the luminosity will be. RSG1L4.5 was one of 3 initial runs with this inner boundary condition; the other two at 20\% and 40\% Temperature increases gained mass too rapidly and never reached a convective steady state.

All our analysis will be done in the convective region starting from  $\approx$ 450$\Rsun$. 
As convection sets in, the luminosity reaches 
$\log(L/L_\odot)=4.5$, with some periodic and stochastic variability which we will discuss in 
more detail in \S\ref{sec:fund3D}. 
Because mass flux through the inner boundary cannot be exactly 0 on a spherical polar mesh even with our reflective velocity boundary condition, a small amount of additional mass enters 
the simulation domain as time goes on. At the end of the simulation, the total mass of this 
model is 16.4$\Msun$. This 6.5\% increase in the total mass of the star ($\approx$20\% in the 
mass inside our simulation domain) is not of concern, as the aim of this work is to create 
realistic 3D envelope models to study the convective structure, not to diagnose a 
mass-luminosity relation in 3D models (which would also be sensitive to core properties). 
The simulation domain for this model is 384 radial zones, with $\delta r/r\approx0.01$, with a free 
outer boundary at $r=22400\Rsun$. 
The choice of a large simulation domain was motivated, in part, to capture any wind structure or 
extended atmosphere, make sure we capture the stellar photosphere so that the outer boundary is 
always in the optically thin limit for the radiation field, as well as to provide ample space 
for expansion in explosions of this envelope model in forthcoming work. With a logarithmic radial 
grid spacing, the large outer boundary is achieved with
small additional cost for our simulation, and $87$ zones lie within $r<1000\Rsun$.

Our second model, referred to hereafter RSG2L4.9, is initialized with the same method as 
RSG1L4.5 for the region that will become convective. This model has the IB at 300$\Rsun$ with 10.79$\Msun$ enclosed, and the total initial mass at 14$M_\odot$.
The simulation domain has 256 radial zones (98 at $r<1000\Rsun$) with $\delta r/r\approx0.01$, with a free outer boundary at $r=6700\Rsun$, still far away from the stellar surface. Between $300\Rsun$ and $400\Rsun$, the initial profile is constructed with the radiative luminosity to be $10^5L_\odot$ and this is kept fixed in the inner boundary (``fixed $L$"). This serves the same purpose as the boundary condition used in the previous model to drive convection for the region above. We therefore similarly only perform our analysis for the region above $\approx$450$\Rsun$. We first run for 740 days with fixed total 
$F_r$ at the inner boundary (including advection and diffusion). After an initial 
relaxation period, this scheme begins to add mass somewhat rapidly, so we switch to fixing only
the diffusive $F_\mathrm{rad,0}$ at the inner boundary. 
This leads to a small, steady decrease in the
total envelope mass from the inner boundary. 
At the end of the simulation, the total mass of 
this model is 12.9$\Msun$. 
In both cases, most of the mass change happens during the initial transient relaxation from the 
initial conditions to a convective structure. From day
4500 to the end of the simulation, the mass inside the
simulation domain changes by less than 1\% ($0.03\Msun$) for RSG1L4.5, and 10\% ($0.2\Msun$) for RSG2L4.9.
The properties of these models are summarized in Table \ref{tab:models}.

\begin{figure*}
\centering
\includegraphics[width=\textwidth]{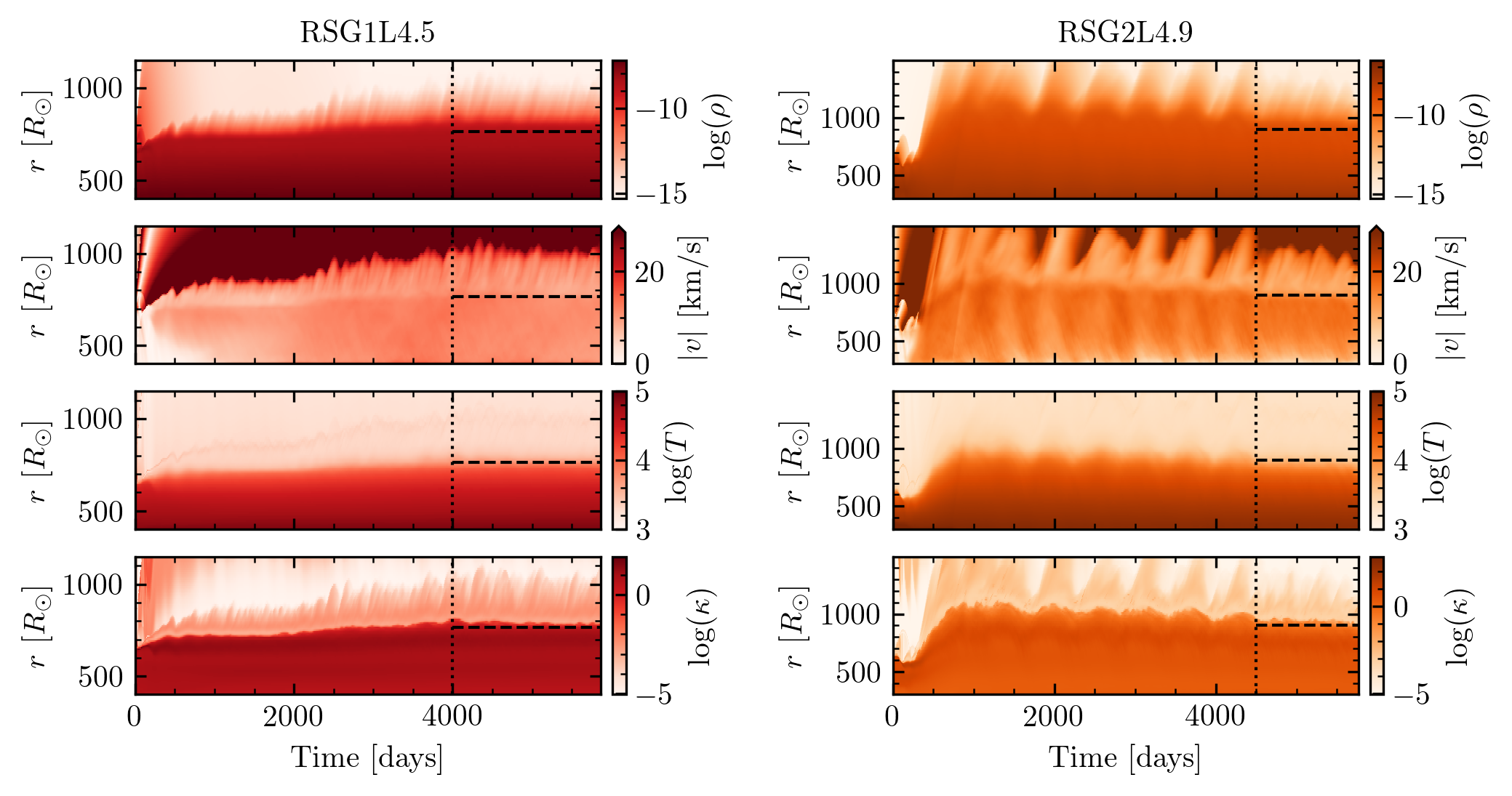} 
\caption{History of the averaged radial profiles for our RSG1L4.5 (red, left) and RSG2L4.9 (orange, right) models. Top to bottom show log(density), convective velocity, log(temperature), and log(opacity). All logarithms are base 10 and units inside the logarithms are cgs, and velocites are reported in km/s. Vertical dotted lines indicate when the envelope appears to have reached a convective steady state. Horizontal dashed lines approximate the region where some fraction of the stellar area has $\tau>\taut$.}
\label{fig:ssprofiles}
\end{figure*}

Radial profiles of both simulations as a function of time are 
shown as space-time diagrams in Fig.~\ref{fig:ssprofiles}. 
Radial density, opacity, and temperature are calculated by finding the
volume-weighted average over spherical shells at each time (which we hereafter denote with $\langle\cdots\rangle$), and the magnitude of the velocity, is calculated from the mass-weighted average over spherical shells (which we hereafter denote with $\langle\cdots\rangle_m$), $|v|=\sqrt{\langle\vr^2+v_\theta^2+v_\phi^2\rangle_m}$. 
Horizontal dashed lines approximate the location where radiation 
begins to dominate the energy transport at late times. 
Inside the dashed line, convection is expected to resemble MLT, with 
denser material sinking as less dense material rises. 
We will explore this expectation in more detail in \S\ref{sec:fund3D} and 
\ref{sec:3Dto1D}. For computational reasons, both models have 
density floors imposed with $\rho_\mathrm{floor}=5.35\times 10^{-16}$ g/cm$^3$. The fast-moving low-density material at very large radii is caused by negligible amounts of density-floor material falling onto the star due to gravity. 

The two simulations begin with an initial transient phase, as convection sets in from the unstable spherically symmetric initial conditions. 
In RSG1L4.5, the intial transient phase is accompanied by some material being launched outwards, falling back onto the stellar surface around day 2500. Additionally, convection begins to appear at the density inversion near the stellar surface, and makes its way to the IB by $\approx1000$ days. By day 2000, the amplitudes of the convective velocities steady and by day 4000 the RSG1L4.5 simulation appears to have fully settled into equilibrium, with regular fluctuations in the stellar properties particularly in the region above $\tau=\taut$. 

In RSG2L4.9, the fixed luminosity at the IB triggers convection at small radii in addition to the surface, so convection sets in quickly. The initial transient causes a sharp increase in the mass contained in the stellar envelope coming from the inner boundary, accompanied by a rapid expansion of the envelope around day 500. With the change in inner boundary condition at day 740, the rapid growth ceases, and by day 2000 the model begins to settle into a pattern of semi-regular oscillations. By day 4500, the amplitude of radial fluctuations dies down and the envelope exhibits similar steady-state behavior to the RSG1L4.5 simulation with larger radial extent and higher velocities.
We now check this apparent steady-state behavior for both simulations. 

\subsection{Defining a `Steady State'}
\begin{figure*}
\centering
\includegraphics[width=0.45\textwidth]{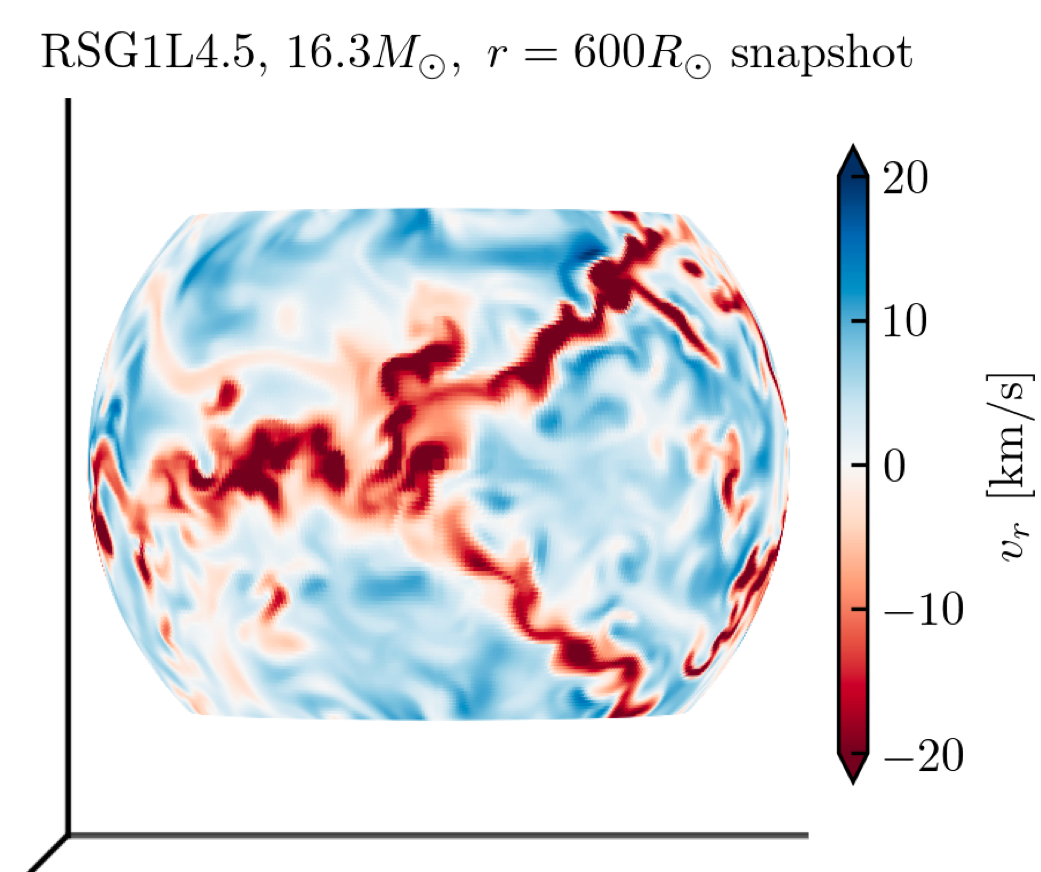}
\hspace{0.15in}
\includegraphics[width=0.45\textwidth]{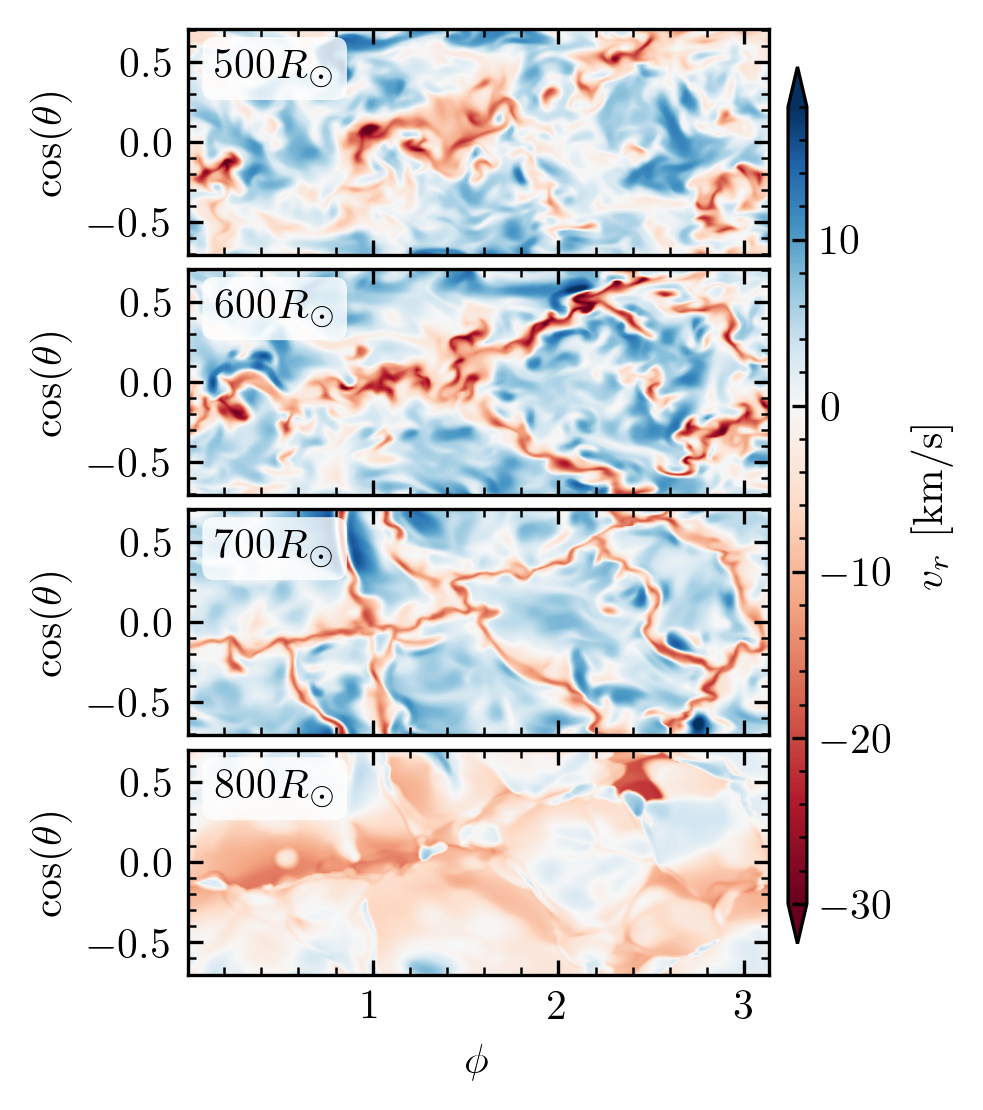}
\caption{Left: Surface rendering of the radial velocity fluctuations at $r=600\Rsun$ in RSG1L4.5 at day 4707. Right: Snapshot of radial velocity slices for the same model. Panels show radial slices at $r=500\Rsun$ (top) to $800\Rsun$ (bottom) in $100\Rsun$ intervals, and axes show the extent in azimuth $\phi$ and co-polar angle cos($\theta$).
The volume-weighted mean sound speed is 33\,km/s at 
$r=500\Rsun$, 26\,km/s at $600\Rsun$, 19\,km/s at $700\Rsun$, and 8\,km/s 
at $800\Rsun$.
}
\label{fig:vradial}
\end{figure*}

By the end of the simulations, both models have thermal and kinetic
energy content $E_\mathrm{fluid}=\int(E+E_r)\mathrm{d}V$ comparable to the 
binding energy $E_\mathrm{bind}=\int(\rho\Phi)\mathrm{d}V$, 
with a ratio of $E_\mathrm{fluid}/|E_\mathrm{bind}|$ of 0.23 for 
RSG1L4.5 (with $E_\mathrm{tot}=E_\mathrm{fluid}+E_\mathrm{bind}=
-3.0\times10^{47}$erg extending the volume to $\Omega=4\pi$) and 0.32
for RSG2L4.9 ($E_\mathrm{tot}=-1.4\times10^{47}$erg). 
The comparable gas and binding energies reinforce our choice of 
including the envelope mass in our gravitational.

\begin{figure}
\centering
\includegraphics[width=\columnwidth]{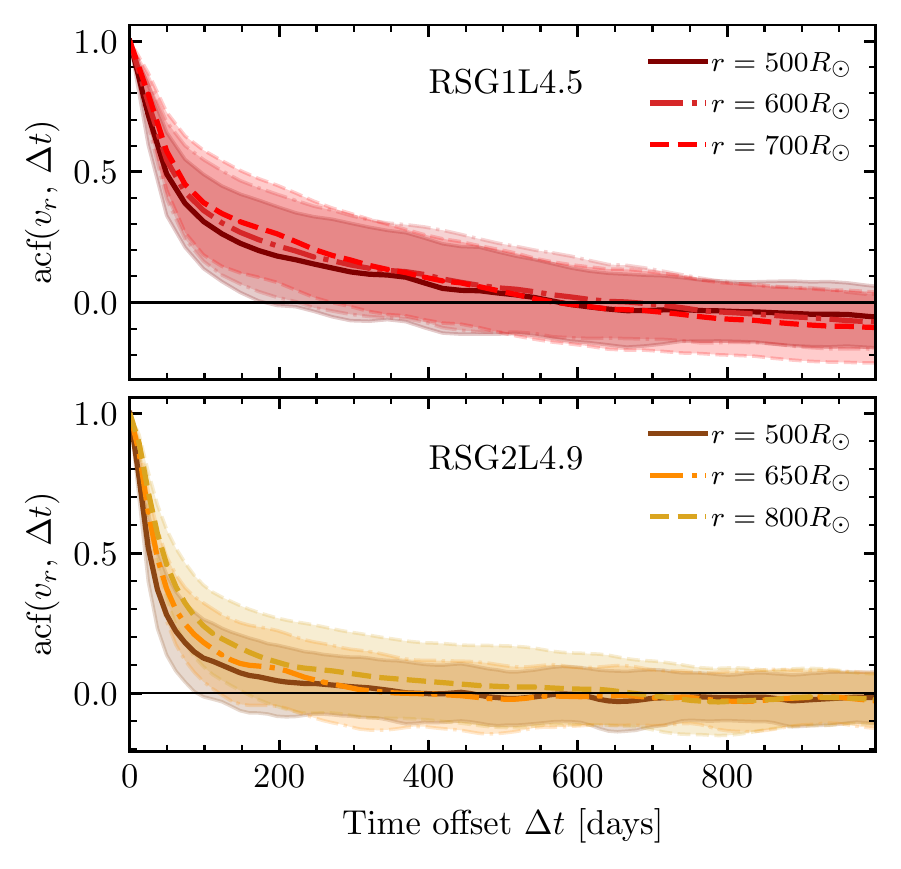} 
\caption{Average autocorrelation of radial velocities across different angles at different radii for the RSG1L4.5 model (upper panel) and the RSG2L4.9 model (lower panel) as a function of the time lag. All data in this plot are after the simulations have run for 1000 days. The shaded region gives the standard deviation of the acf across the different angles.}
\label{fig:sscorr}
\end{figure}

The convective plumes show large ($\gtapprox200\Rsun$) vertical and lateral extents, leading to a nearly-radius-independent velocity profile with an-order-of-magnitude scatter, shown in Fig.~\ref{fig:vradial}. 
When fluid flow is this coherent, the velocity field
will be time-correlated for around an eddy-turnover time at any given spatial location. Beyond this timescale, we expect no memory of past convective plumes at a fixed coordinate. To start our exploration of the timescale required for the model to reach equilibrium, we calculate the autocorrelation of the radial velocity at fixed coordinates. The autocorrelation function for an arbitrary time-dependent parameter $Y(t)$ is defined for time lag $\Delta t$ by 
\begin{equation}
    \mathrm{acf}(Y,\Delta t)=\frac{\int{(Y(t)-\Bar{Y})(Y(t+\Delta t)-\Bar{Y})\intdt}}{\int(Y(t)-\Bar{Y})^2\intdt},
\end{equation}
where $\Bar{Y}$ is the time-averaged value. Fig.~\ref{fig:sscorr} shows the autocorrelation 
function for the radial velocity, acf($\vr, \Delta t$), for a few different radii in each model. Dark lines give the mean of the autocorrelation functions at 169 angles distributed 
across the stellar model, and the shaded areas give the standard deviation of the acf at each 
radius. Only radial velocities after day 1000 are considered for each model.
The less luminous RSG1L4.5 model stays correlated for 
$\approx$550 days, whereas the more luminous RSG2L4.9 model decorrelates faster, over a 
timescale of $\approx$300 days. This is because the more luminous model exhibits larger
convective velocities as required to carry the flux, even though the mass and radii are comparable. These timescales are short compared to the simulation duration, so we can proceed in our analysis with additional confidence that both models have reached their convective steady state after $\approx$4000 days.

\begin{figure}
\centering
\includegraphics[width=\columnwidth]{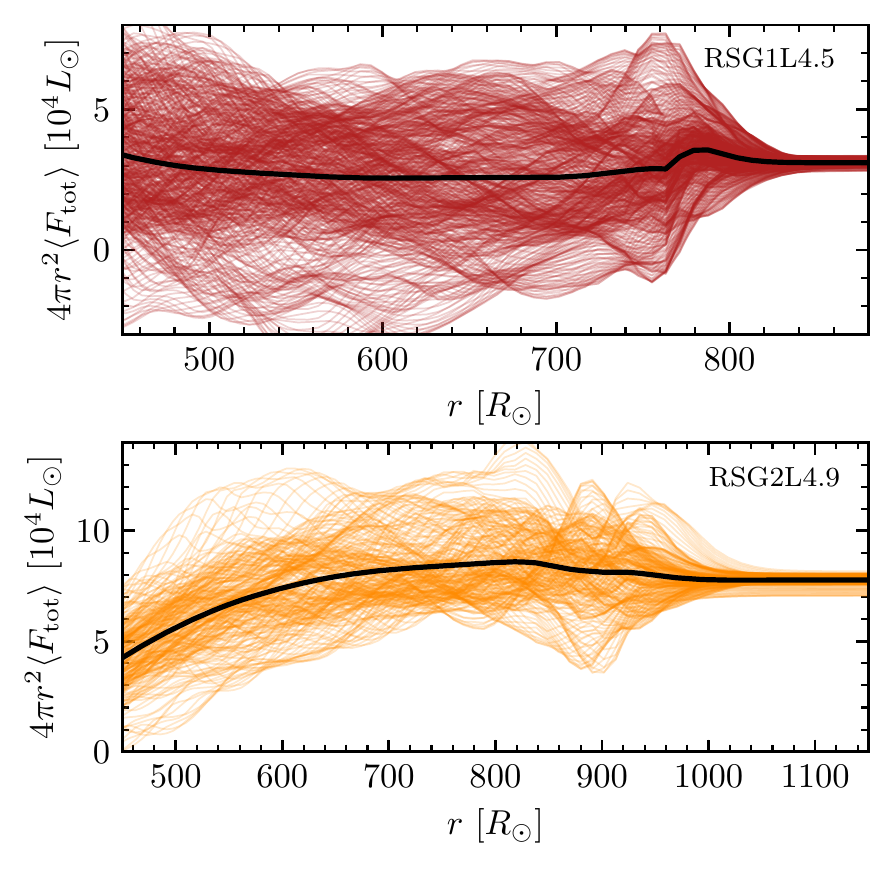} 
\caption{Volume-average of $4\pi r^2((E+\Pgas)\vr+F_{\mathrm{r},\hat{r}}-(\rho\vr\Phi))$ at 
different times (thin colored lines) compared to the time-average (solid black line) for our 
RSG1L4.5 (red, upper panel) and RSG2L4.9 (orange, lower panel) models outside $r=450\Rsun$.}
\label{fig:ssflux}
\end{figure}

A direct check as to whether the models have reached a convective steady
state is to explore the RHD equations for the time-averaged profiles. 
The momentum equation quickly equilibrates such that $\partial\rho\bv/\partial t\rightarrow0$ when taking the time-average on a dynamical timescale (i.e. the sound-crossing time across a pressure scale height, 10s to 100s of days in the outer envelope), but the energy equation will only reach equilibrium in our region of interest as 
convection is able to distribute the luminosity over a few eddy turnover
times, which is significantly longer. Combining and rearranging RHD Eqs.~\eqref{eq:RHD} including the source term
\begin{equation}
 G^0_r=\frac{\partial E_{\mathrm{r}}}{\partial t}+\nabla\cdot \boldsymbol{F}_{\mathrm{r}},
\end{equation}
where $\bv{F}_r$ is the total radiation flux (including radiative diffusion $F_\mathrm{rad,0}=-1/3(c/\kappa\rho)(\intd a_rT_r^4/\intd r)$ 
and advection $F_\mathrm{adv}=E_r\vr$), with $g=-\nabla\Phi\cdot\hat{r}$, for spherically symmetric $\Phi(r)$, we recover
 \begin{equation}
\begin{aligned}
&\frac{\partial}{\partial t}(E+E_\mathrm{r})+\nabla\cdot[(E+\Pgas)\boldsymbol{v}+\boldsymbol{F}_\mathrm{r}-(\rho\boldsymbol{v}\Phi)] \\&=-\Phi\nabla\cdot(\rho\boldsymbol{v})=-\Phi\frac{\partial}{\partial t}\rho.
\end{aligned}
\end{equation}
In a steady state, $\partial/\partial t\rightarrow0$ when we take the time average $\langle\cdots\rangle_t$. Taking the radial component of the divergence we find
\begin{equation}
\begin{aligned}
\left\langle\frac{1}{r^2}\frac{\partial}{\partial r}[r^2(E+\Pgas)\vr+r^2 F_{\mathrm{r},\hat{r}}-r^2(\rho\vr\Phi)]\right\rangle_t &=0. \\
\end{aligned}
\end{equation}
Thus if
$r^2\left\langle (E+\Pgas)\vr+F_{\mathrm{r},\hat{r}}-(\rho\vr\Phi)\right\rangle_t\equiv r^2\langle F_\mathrm{tot}\rangle$ is spatially constant, the model can be considered to have equilibrated. This expression
is equivalent to the time-average of the volume-weighted average of the total luminosity $L_\mathrm{tot}$, including enthalpy, gravity, kinetic energy, and 
radiation terms, divided by 4$\pi$. Though there is a net change in  $L_\mathrm{tot}$ due to mass gained/lost near the IB, in the region of interest, our steady state criteria are sufficiently satisfied by both models in the region of interest, as shown in Fig.~\ref{fig:ssflux}. 
The transparent colored lines show the volume-averaged total luminosity from days 
4001$-$5864 in approximately 3 day intervals for the RSG1L4.5 model (red) and evolution from 
day 4501$-$5766 in approximately 5 day intervals for the RSG2L4.9 model (orange). The solid black lines give the time-average of this quantity, whose variance at different radii is significantly less than the scatter at different times. The small number of distinct convective plumes implies a fundamental variance in the stellar luminosity reflected in the scatter at large radii that we discuss in more detail in \S \ref{sec:fund3D}.

\section{3D Model Properties}\label{sec:3Dprops}

Having shown how to initiatlize a 3D RSG model and evolve it to its effectively equilibrium state, we now will describe the properties
of the resulting models, compare to prior work, and discuss some of the unique properties of these highly luminous models.
 
\subsection{Convective Properties and Comparison to Prior 3D \COBOLD\ RSG Work}
\label{sec:compare3D}

\begin{figure}
\centering
\includegraphics[width=\columnwidth]{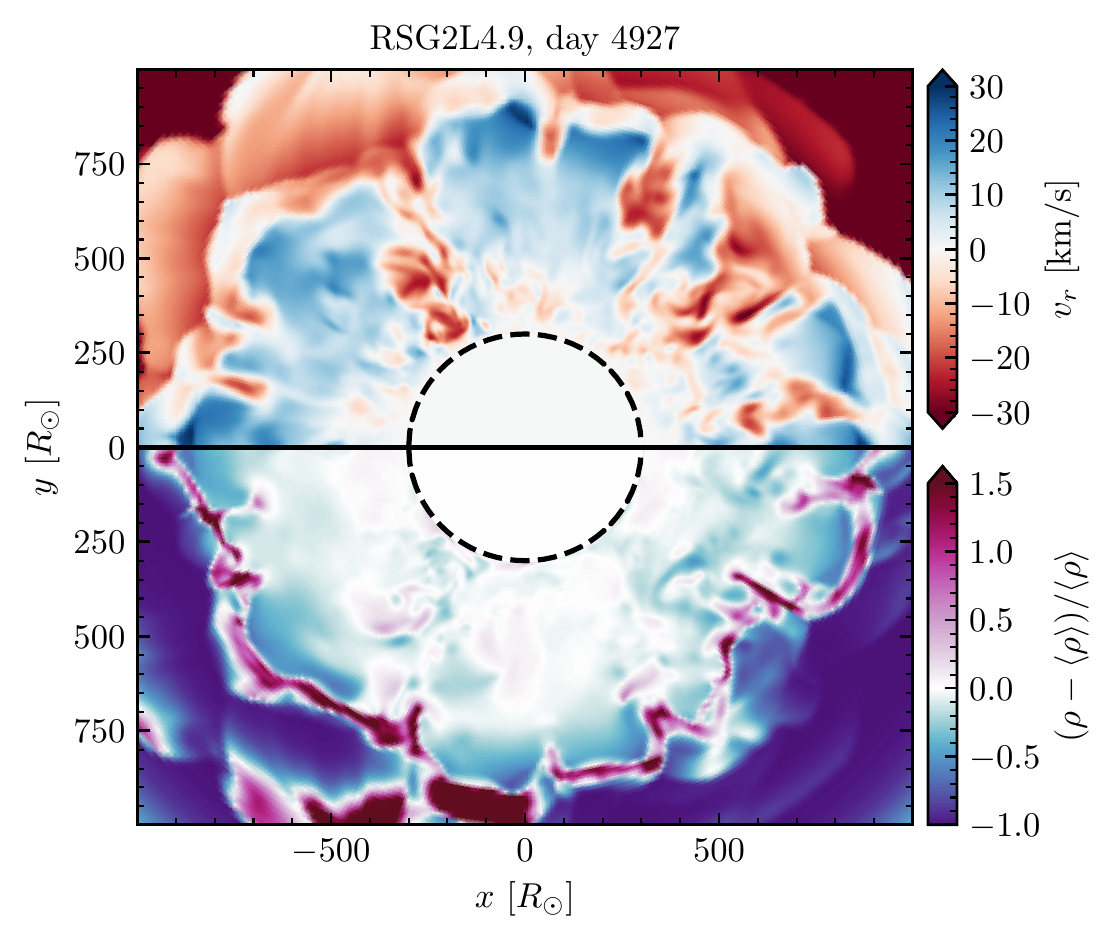}
\caption{ \label{fig:eyecandy} Equatorial ($\theta=\pi/2, z=0$) slice of a characteristic snapshot of the RSG2L4.9 model, showing (upper half of the figure:) radial velocities and (lower half of the figure:) density fluctuations relative to the shellular volume-averaged density at each radius. The simulation domain is from $\phi=0$ to $\pi$; thus the image is reflected about $y=0$ as indicated by the axis labels.}
\end{figure}

\begin{figure*}
\centering
\includegraphics[width=0.48\textwidth]{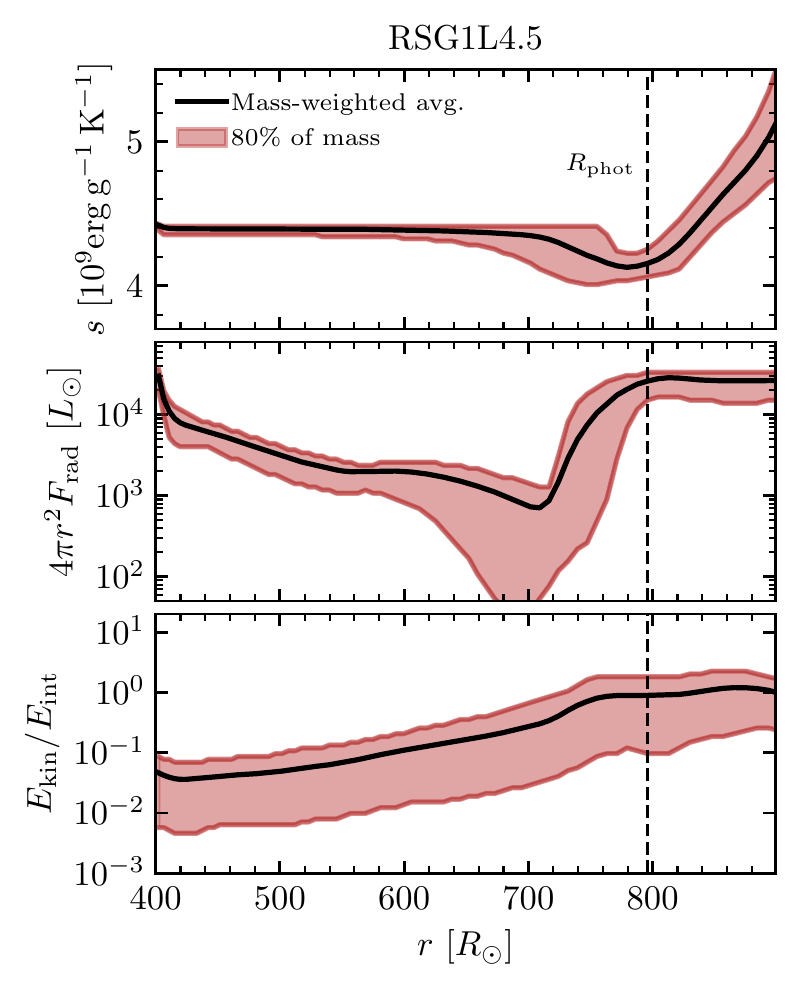}
\includegraphics[width=0.48\textwidth]{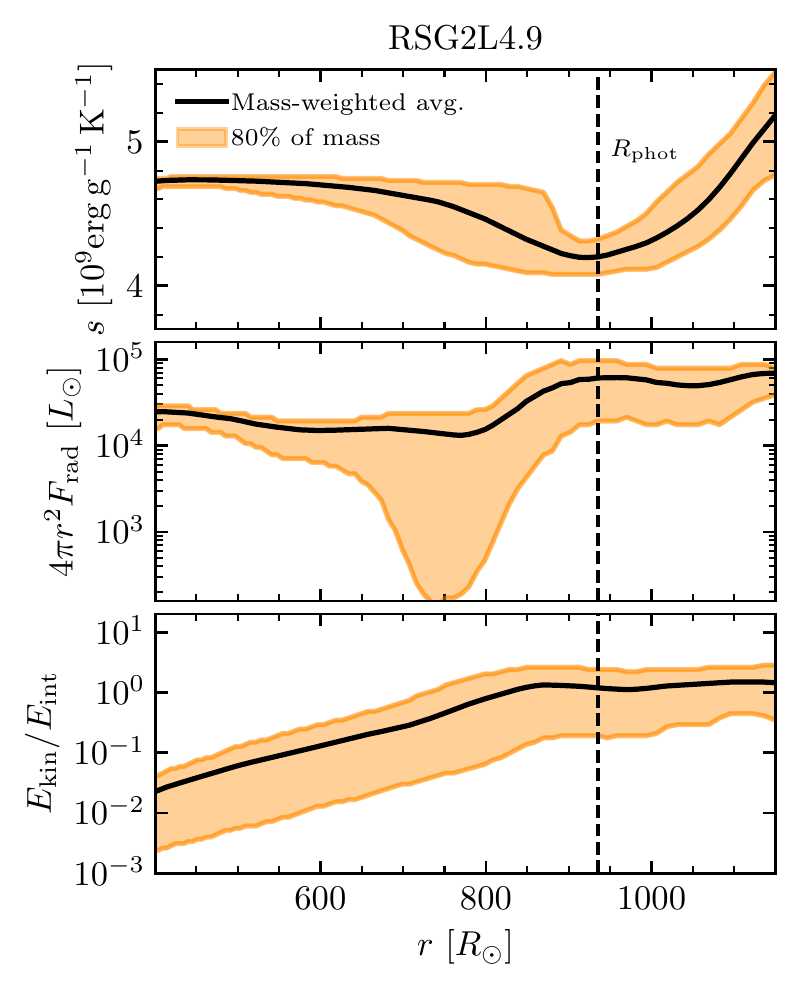}
\caption{ \label{fig:SLPratio} Specific entropy (top panels), radiative luminosity (middle panels), and ratio of the turbulent kinetic energy to the internal energy (bottom panels) in our RSG1L4.5 model at day 4707 (red, left) and RSG2L4.9 at day 4927 (orange, right). Mass-weighted averages are shown in black, with 80\% of the mass lying within the shaded regions. The 1D photosphere, where $\langle{L(r)}\rangle=4\pi r^2\sigma_\mathrm{SB}\langle{T_r(r)}\rangle^4$, is given by the vertical dashed line.}
\end{figure*}

\begin{figure*}
\centering
\includegraphics[width=0.48\textwidth]{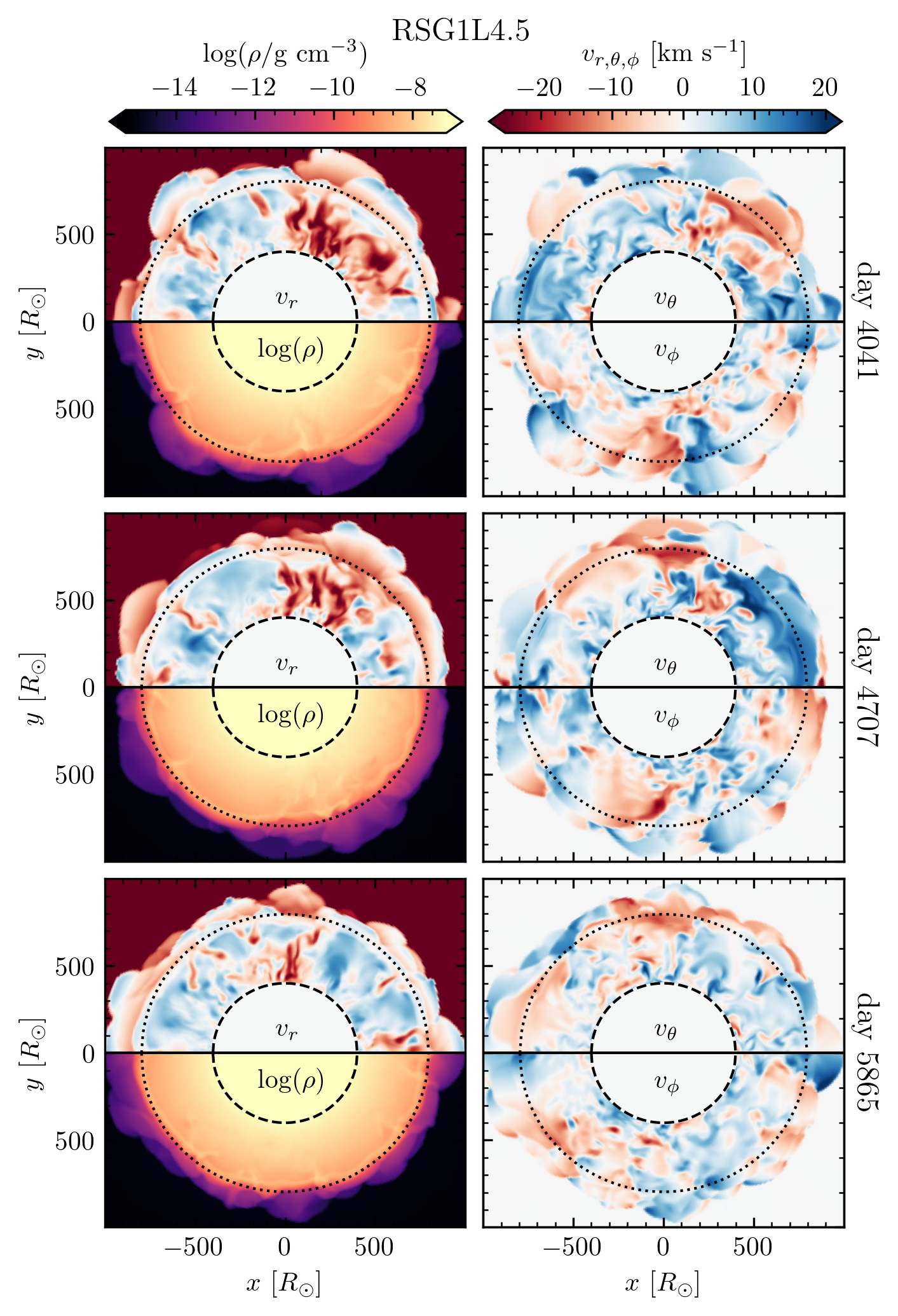}
\includegraphics[width=0.48\textwidth]{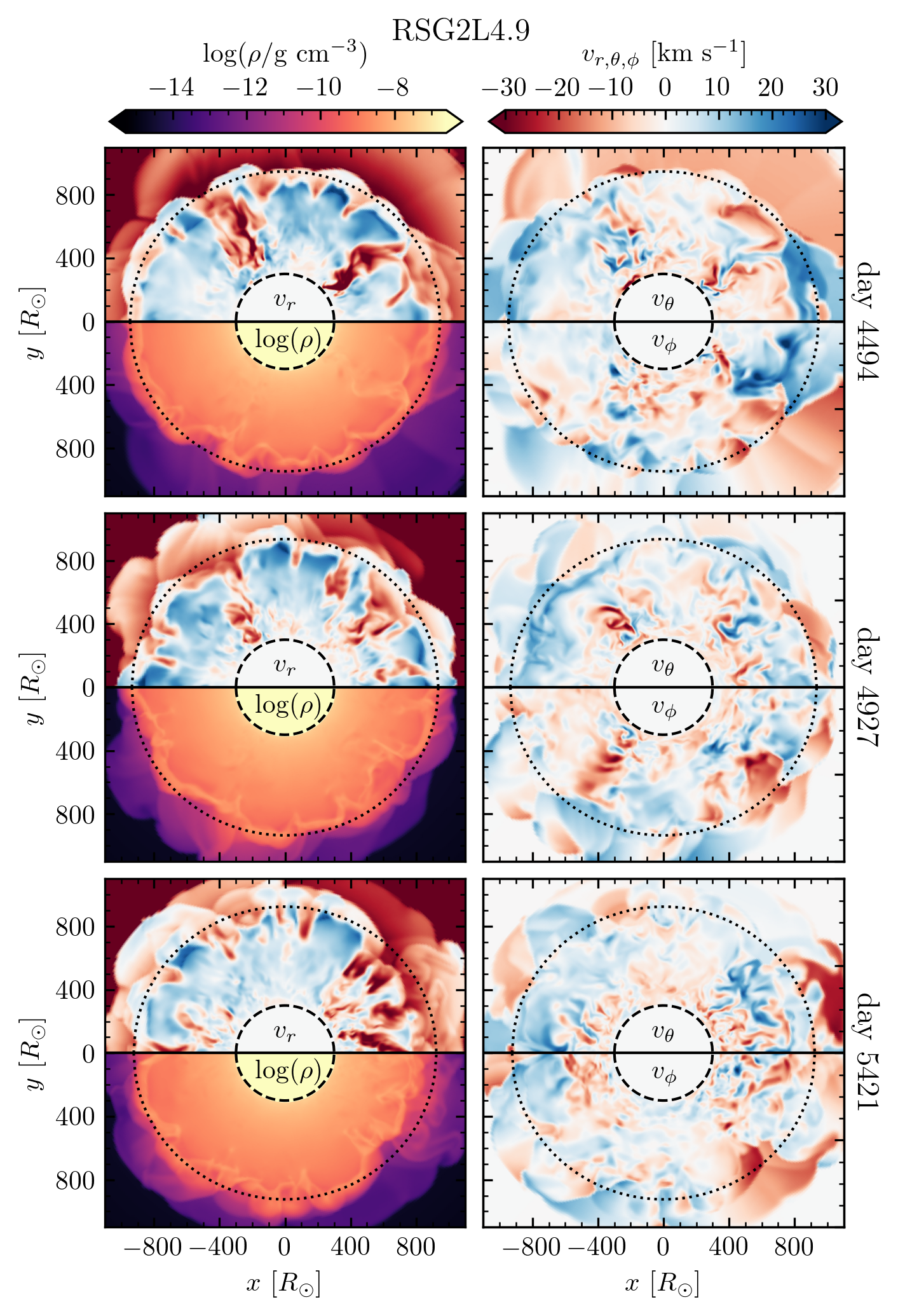}
\caption{Equatorial ($\theta=\pi/2, z=0$) slices for our RSG1L4.5 (left 6 panels) and RSG2L4.9 (right 6 panels) models at different simulation times. At each time for each model, the left panel shows radial velocity (red/blue colors) and density (orange/purple colors) is beneath. The right panel shows polar tangential velocity ($v_\theta$) and azimuthal velocity ($v_\phi$) is beneath.  The simulation domain is from $\phi=0$ to $\pi$; thus the y-axis is reflected in the lower half of each panel as indicated by axis labels. With $\theta=0$ along the $+z$ axis and $\phi=0$ along the $+x$ axis, $v_\theta>0$ indicates material flowing into the page, and $v_\phi>0$ indicates material flowing clockwise around the page (due to the inverted y-axis). The dashed line indicates the simulation inner boundary and the dotted line denotes the 1D photosphere.}
\label{fig:pokeballs}
\end{figure*}

Aspects of the observable 3D structure of RSGs have been studied in a series of pioneering papers (e.g. \cite{Chiavassa2009,Chiavassa2010,Chiavassa2011a,Chiavassa2011b,Kravchenko2018}) using the RHD ``star-in-a-box'' COnservative COde for the COmputation of COmpressible COnvection in a BOx of L Dimensions, L=2,3 
\citep[\COBOLD,][]{Freytag2002,Freytag2010,Freytag2012}.
In those simulations, the computational grid was cubic equidistant with typical mesh spacing of $\approx8.6\Rsun$, with LTE radiation transport by short characteristics using opacity tables as function of $P,\ T$ interpolated from PHOENIX data at $T\ltapprox12,000\,$K \citep{Hauschildt1997} and OPAL values \citep{Iglesias1992} at higher $T$. The EOS included ideal gas and ionization, but radiation was only present in the energy equation, and not as a pressure source. The gravitational potential was modeled by a Plummer potential $\Phi=-GM_{*}/(r_{0}^{4}+r^{4}/\sqrt{1+(r/r_{1})^{8}})^{1/4}$ fixed to the static Cartesian mesh with $M_*=12\Msun$ and $3\Msun$ of material contained in the simulation domain, and the luminosity was supplied via an energy source within the inner Plummer radius ($r_0$).

That work focused on stellar properties at low optical 
depth, where radiation transport dominates, and have been compared to recent tomagraphic 
observations of nearby RSGs to interpret their surface convective structure \citep[e.g.,][]{Kravchenko2019,Kravchenko2020, Kravchenko2021}. 
The neglect of radiation pressure deep within
the star inhibited the ability to correctly simulate the deeper 
nearly-constant-entropy convective zone there. Hence, the convective flux
in the interior of those RSG models is 
significantly lower than the radiative flux, with radiation 
carrying over 80\% of the flux everywhere.
Because of this, those simulations exhibit a positive entropy gradient out 
to $\approx$75\% of the stellar radius (see Fig. 3 of \citealt{Chiavassa2011b}). 
While this is no concern when restricting analysis to the 
turbulent surface layers where the entropy profile does decline, 
it is counter to the theoretical expectations for a fully 
turbulent RSG envelope, which should have a 
nearly-flat, declining entropy profile throughout the convective 
envelope, with enthalpy and kinetic flux accounting for 
a significant fraction of the total flux.

In agreement with the \COBOLD\ models, our \Athena\ RSG simulations
show a handful of large-scale, coherent convective plumes across the star, with radial velocities of tens of km/s and density fluctuations of 10\% increasing to factors of a few at larger radii (see Figs.~\ref{fig:vradial} and \ref{fig:eyecandy}). 
As emphasized by \citet{Stein1998}, we see a topology of large area upwellings surrounded by narrow lanes of downward flows.
Additionally, the specific entropy, radiative luminosity, and ratio of kinetic to thermal energy density in representative snapshots of our two models are shown in Fig.~\ref{fig:SLPratio}. The red/orange shaded regions give a sense of the scatter. Like in the \COBOLD\ models, we observe a `halo' of bound, high-entropy material above the conventional photosphere, with density fluctuations exceeding an order of magnitude in the outer-most parts of the star. The entropy profile in the interior of our \Athena\ models is nearly adiabatic,  
declining slightly out to $100-200 R_\odot$ beneath the 1D photosphere, and declining more rapidly as radiation is able to carry more of the flux. These entropy profiles are similar to those seen in lower-luminosity RHD models \citep[e.g.][]{Stein1998,Magic2015}.

Following \citet{Chiavassa2011b}, we define the 1D 
photosphere by calculating 1D radial profiles of the 
luminosity $\langle L(r)\rangle$ and radiation temperature $\langle 
T_r(r)\rangle$; $r=\rphot$ is then defined as the 
location where $\langle L(r)\rangle=4\pi r^2\sigma_\mathrm{SB}\langle T_r(r)\rangle^4$. The energy transport
in the stellar interior is dominated by convection, and
radiation carries $\approx10\%$ of the luminosity in the 
convective region. Moreover, the turbulent kinetic energy density from the vigorous convective motions dominates over the thermal energy in the outer envelope, in agreement with the findings of \citep{Chiavassa2011b}.

\subsection{Stochastic Angular Momentum}\label{sec:angularmomentum}

The 3D properties of convection in RSG interiors are also of interest for predicting properties of the remnant in failed SNe \citep[e.g.][]{Coughlin2018, Quataert2019}. 
Recently, \citet{Antoni2021} completed a detailed study of convective fluid motion
with applications to collapsing RSGs using 3D hydrodynamical simulations of 
idealized RSG models spanning a factor of 20 in stellar radius. Their work 
considers an ideal gas with polytropic index $\gamma=1.462$ in a Plummer potential 
$\Phi=-GM/(r^n+a^n)^{1/n}$  with $n=8$ for a smoothing radius $a\ll r_o$ in
their dimensionless code units where $r_o=\rphot/6$. This converges to a point 
mass in their region of interest. These pure-hydro simulations enforce a photospheric 
radius by providing a cooling sink at large fixed radii and smoothly decreasing 
the temperature outside the photosphere to be equal to their temperature floor. 
Their study focused on quantifying the randomly distributed angular momentum of the 
inner layers of the convective RSG envelope, and how these shells evolve during later collapse. 

In our models, we likewise observe large tangential velocity fluctuations due to the random convective fluid motion with coherence across many scale heights.
Fig.~\ref{fig:pokeballs} shows the radial and tangential 
components of the fluid velocity for equatorial ($z=0$, $\theta=\pi/2$) slices 
through our \Athena\ models, as well as the density structure. The large radial 
velocity plumes carry material out beyond $\rphot$ (the dotted lines). As the fluid becomes optically thin, the 
temperature plummets and the pressure scale height drops, and the very
large convective plumes fragment into smaller bubbles of surface convection. This is especially apparent in the more luminous RSG2L4.9 model (right 6 panels). 
Additionally, there is some outward-moving high-density material 
evident at large radii. We will discuss this material in greater 
detail in \S \ref{sec:fund3D}. The velocities in $r,\ \theta,$ and $\phi$ are 
comparable, with values of tens of km/s. The 
tangential flows ($v_\theta$ and $v_\phi$) exhibit smaller-scale 
structures at smaller radii, in agreement with the results of \citet{Antoni2021}. 

\begin{figure}
\centering
\includegraphics[width=\columnwidth]{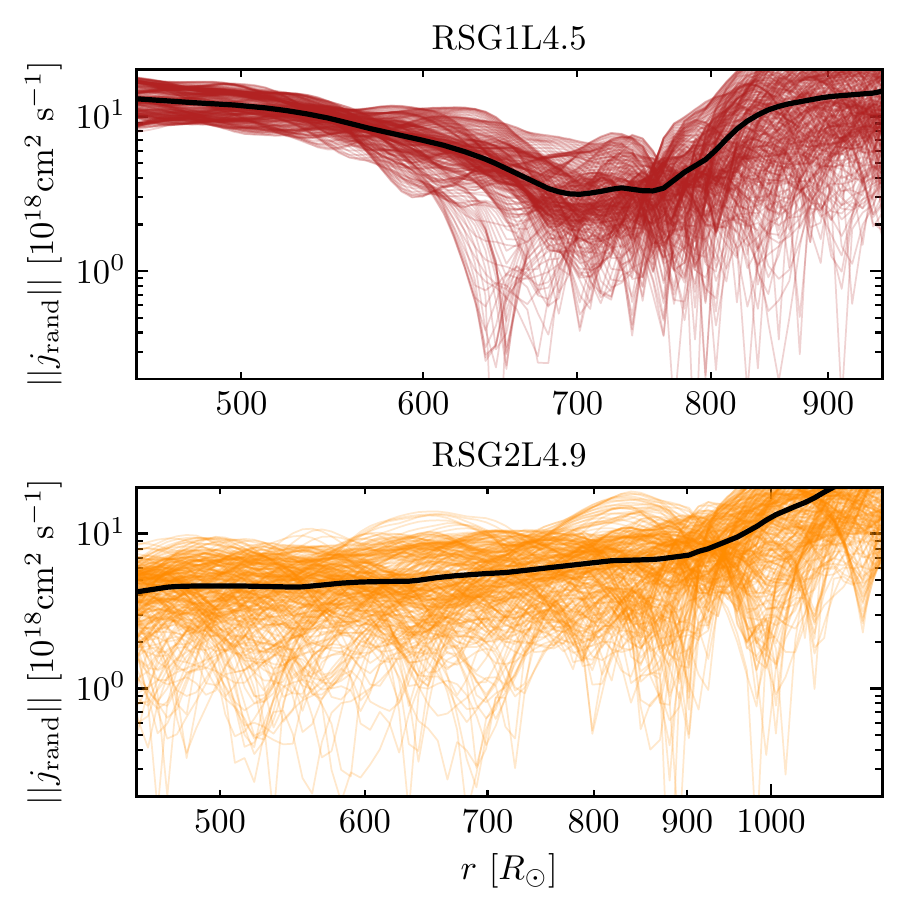}
\caption{\label{fig:jrand}Specific angular momentum profiles at different snapshots (thin colored lines), compared to the time-average (solid black lines), for our RSG1L4.5 (red, upper panel) and RSG2L4.9 models (orange, lower panel).}
\end{figure}

Although the net angular momentum in the envelope is nearly zero, 
these tangential velocity fluctuations result in finite specific 
angular momentum $j_\mathrm{rand}$ at a given radius at any given time.
The magnitude (denoted $||\cdots||$) of the mass-weighted average of the
random specific angular momentum profiles, equivalent to 
$||j_\mathrm{rand}||=||\langle\rho\bv\times\br\rangle/\langle\rho\rangle||$, is shown in Fig. \ref{fig:jrand}. As in Fig. 
\ref{fig:ssflux}, faint colored lines correspond to individual snapshots in our 
models, and the solid black line indicates the time-average. In agreement with 
\citet{Antoni2021}, these simulations exhibit relatively flat specific
angular momentum profiles, pointing to the non-local coherent nature 
of the convective plumes. 
Due to the high $\sim10$ km/s tangential 
velocities and the fact that our simulation domain 
emphasizes large radii ($r>400\Rsun$), we find specific 
angular momenta of $10^{18}-2\times10^{19}$cm$^2$/s
throughout our simulation domain. Transforming to a local rotational velocity $\omega_\mathrm{rot}=||j_\mathrm{rand}||/r^2$, this corresponds to a range of $\omega_\mathrm{rot}$, declining from $\approx$10$^{-3}$ rad/day to 10$^{-4}$ rad/day before rising outside $\rphot$ in RSG1L4.5, and a flatter, slightly declining time-averaged profile around a few $\times10^{-3}$ rad/day in RSG2L4.9 with the scatter between snapshots ranging from a few times $10^{-5}$ to 8$\times 10^{-4}$ rad/day.
These values are slightly larger than those reported by \citet{Antoni2021}, likely
owing to the larger convective velocities present in our simulations.\footnote{\citet{Antoni2021} reported values for volume-averaged specific angular momentum 
$||\langle\bv\times\br\rangle||$, which are nearly equivalent to the mass-weighted average in their region of interest where $r<\frac{5}{6}\rphot$. At large radii, the mass-weighted average, equal to the total angular momentum in a shell divided by the total mass of the shell, favors the denser turbulent plumes rather than the high-volume low-density background, leading to larger values of $||j_\mathrm{rand}||$. However this effect is not so dramatic that reporting the volume-weighted average would account for the apparent difference.} 
Following \citet{Quataert2019}, we should reduce our estimate by the expected scaling for the larger number of eddies available in a $4\pi$ steradian simulation, 
which would then be a factor of $(\Omega/4\pi)^{1/2}\approx0.6\times$ smaller. This modifies our values to a few$\times 10^{17}-10^{19}$ cm$^2$/s, which are closer to those found by \citet{Antoni2021}.

\subsection{Nature of 3D Convective Structure}\label{sec:fluidstructure}

\begin{figure}
\centering
\includegraphics[width=\columnwidth]{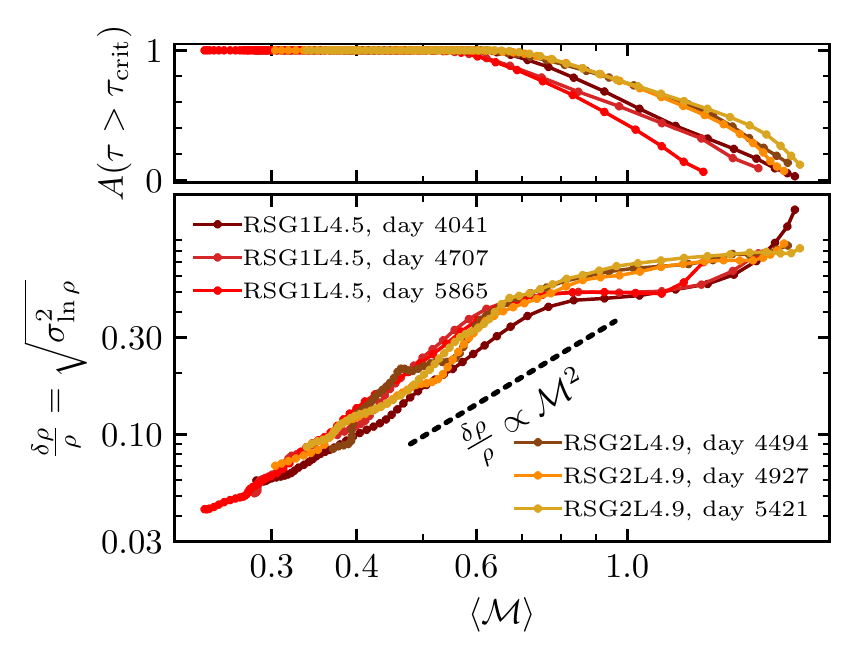}
\caption{\label{fig:rhovmach} Characteristic density fluctuations versus the 
(volume-weighted) average turbulent mach number for the 6 model snapshots shown in Fig.~\ref{fig:pokeballs}. Each point corresponds to the averaged 
value in each radial shell. The upper panel shows the area fraction of the 
star at each location with $\tau<\taut$ along a radial line of sight. 
Where $A(\tau>\taut)=1$ we expect the flow to follow the 
$\delta\rho/\rho\propto\mathcal{M}^2$ scaling, indicated by the black dashed line. }
\end{figure}

In a clumpy or turbulent medium, density fluctuations are often characterized by $\sigma_\rho^2=\frac{\langle\rho^2\rangle}{\langle\rho\rangle^2}$ \citep[see, e.g.][in the context of stellar winds]{Owocki2018}. For a log-normal density distribution typical of convective flows, this 
is related to the characteristic density fluctuations by $(\delta\rho/\rho)^2=\sigma_{\ln\rho}^2=\ln\left(\frac{\langle\rho^2\rangle}{\langle\rho\rangle^2}\right)$ \citep{Schultz2020}. 
Locally, the buoyant acceleration felt by a perturbed 
fluid element with density $\rho+\delta\rho$ will be related to gravity as $a\approx(\delta\rho/\rho)g$. 
The perturbation will approximately traverse a scale height (or mixing length) in time $t$ with velocity $v\sim~a~t\sim(\delta\rho/\rho)(H/v)g$. Thus $v^2\sim~gH(\delta\rho/\rho)$,
or $\delta\rho/\rho\propto\mathcal{M}^2$, where $\mathcal{M}=v/c_\mathrm{s}$ is the 
Mach number. 
Fig.~\ref{fig:rhovmach} shows the characteristic density fluctuations 
$\delta\rho/\rho=\sqrt{\sigma^2_{\ln\rho}}$ versus the average Mach 
number in each spherical shell for $450\Rsun<r<\rphot$ using the snapshots shown in Fig.~\ref{fig:pokeballs}.  
The area fraction of the star where the optical depth along a radial line of sight is greater than the angle-averaged $\taut$, $A(\tau>\taut)$, is also shown. Where $A(\tau>\taut)$ is large, the fluid in both models
follows closely with the expected 
$\delta\rho/\rho\propto\mathcal{M}^2$ scaling. As 
$A(\tau>\taut)$ decreases and convection no longer dominates the energy 
transport everywhere, the scaling flattens and the density 
fluctuations begin to saturate. Other snapshots exhibit the same behavior in both models.  

\begin{figure*}
\centering
\includegraphics[width=0.95\textwidth]{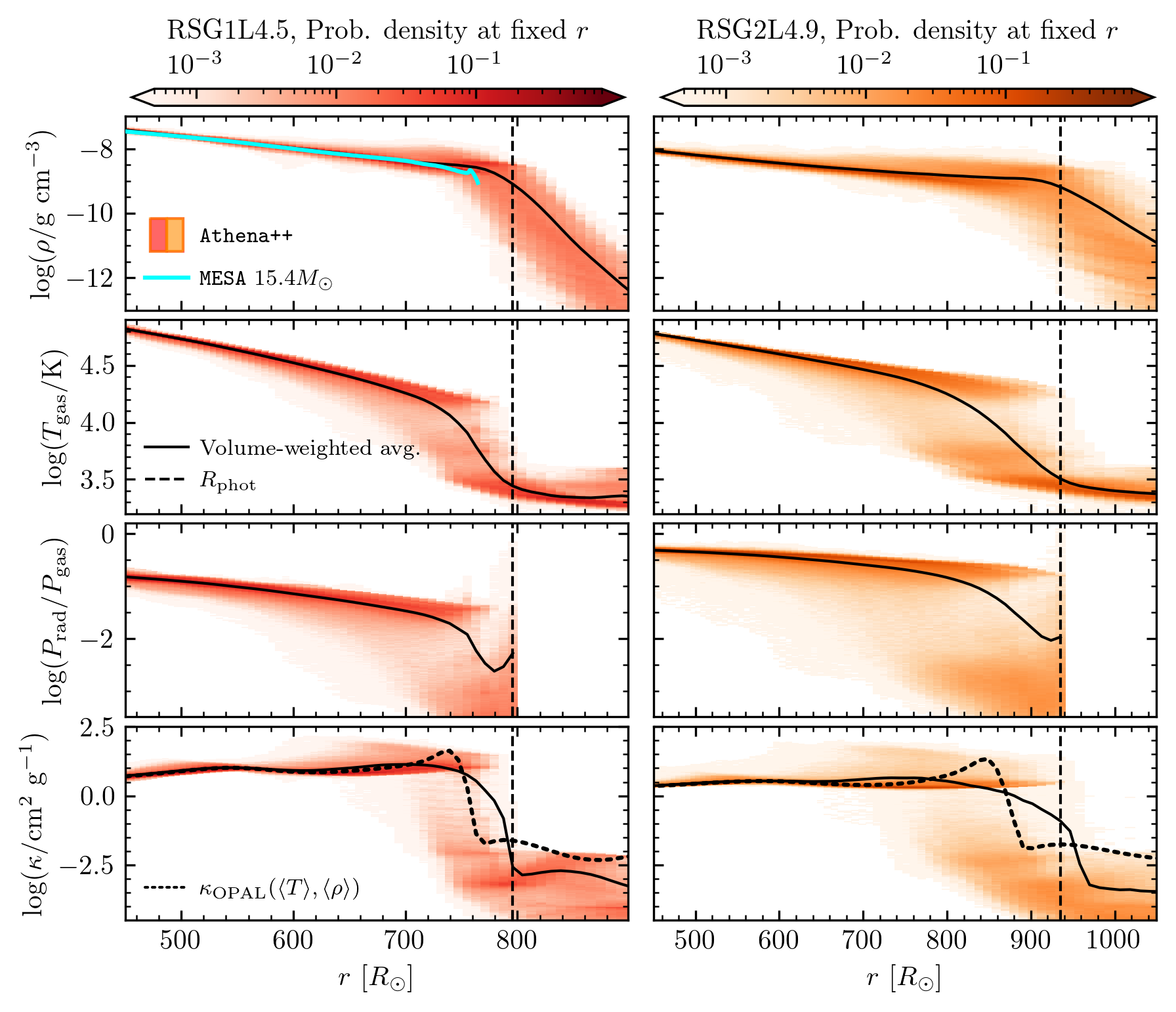}
\caption{Top to bottom: Density, temperature, 
$\Prad/\Pgas$ ratio, and opacity for our RSG1L4.5 model at day 4707 
(red, left panels) and our RSG2L4.9 model at day 4927 (orange, right 
panels). Color saturation indicates the volume-weighted probability of 
finding a fluid element at a given ($\rho, T, H/r, \Prad/\Pgas, \kappa$)
at each radial coordinate, and solid black lines give
the volume-weighted averages of each (non-log) quantity ($\langle\rho\rangle, \langle T\rangle, \langle H/r\rangle, \langle\Prad/\Pgas\rangle, \langle\kappa\rangle$).
The vertical black dashed line is $\rphot$.  As $\Prad$ is not defined 
in the free-streaming regime, the ratio $\Prad/\Pgas$ is only shown for 
$r\leq\rphot$. The $\kappa$ panels (bottom) show both the volume-averaged 
opacity $\log(\langle\kappa\rangle)$ reported by \Athena\ (solid lines), and the 
recovered OPAL opacity (dotted lines) from the volume-averaged $T$ and $\rho$ profiles. For reference, the cyan line in the upper left panel shows the density profile of the fiducial \MESA\ model (green line in the left panels of Fig.~\ref{fig:panels1}).}
\label{fig:3Dpanels}
\end{figure*}

Other stellar properties also exhibit large fluctuations at a given radius, particularly in the
outer stellar layers, where the transition to radiation-dominated 
energy transport does not happen at one single radial location. 
Fig.~\ref{fig:3Dpanels} shows radial profiles of the density, gas 
temperature, $\Prad/\Pgas$ ratio, and opacity 
for characteristic snapshots of RSG1L4.5 and RSG2L4.9 (day 4707 and 
4927, respectively), with solid black lines showing the 
volume-averaged radial profiles and color indicating the scatter. The 
density, which falls like $1/r^2$ in the nearly-adiabatic interior, 
exhibits variations over 2-3 orders of magnitude near $\rphot$, with an extended atmosphere which is absent in 1D models (compare to the cyan line in the upper left panel).

The ratio of radiation to gas pressure is also significant, with 
$\langle\Prad/\Pgas\rangle=0.15$ at $r=450\Rsun$ 
in the RSG1L4.5 model, and $\langle\Prad/\Pgas\rangle=0.48$ at $r=450\Rsun$ in
RSG2L4.9 owing to the larger luminosity and lower density within the 
envelope. Moreover, rather than a smooth transition from the H opacity
peak to electron scattering, the temperature and opacity display 
bimodal behavior in the $\approx100-200\Rsun$ region beneath $\rphot$.
This bimodality is not seen in the $\rho$ profile.
Near and even within the 1D photosphere, at some angular locations 
outer convective plumes exhibit large opacities, whereas other angular
locations are dominated by cool material beyond the H opacity peak. Due to the dramatic 4-order-of-magnitude differences in opacity of 
different material at fixed radius, a linear volume-average of the 
opacity, given by the black line in the bottom panels, will 
necessarily favor the high-opacity material. 
Most notably, the opacity above which $L$ locally exceeds $\Ledd$, 
$\kappa_\mathrm{Edd}=4\pi Gc m/\Lsurf$, is $\kedd\approx6$ cm$^2$/g for RSG1L4.5 
and $\kedd\approx2$ cm$^2$/g for RSG2L4.9. 
In the bimodal transitionary region, a large fraction of the material has $\kappa>\kedd$!
Moreover, the presence of bimodal temperature and opacity distributions in this transitionary
region causes a smearing out of the H-opacity peak, so the H opacity 
cliff, predicted by OPAL using the volume-averaged $\rho$ and $T$ 
profiles, is less steep in the 3D simulation. 

This transitionary region corresponds to the place where 
$A(\tau>\taut)$ goes from 1 to 0 and the turbulent motions deviate 
from classical convection. We now turn to exploring the fundamentally 
3D properties of this convection in the RSG regime. 

\subsection{The Transition to Radiation-Dominated Energy Transport}
\label{sec:fund3D}

In classical MLT, a flow of fluid parcels, or ``bubbles," approximately 
maintain their entropy and carry heat out as they rise with convective 
velocity $\vc$ over a mixing length $\ell$ (See \citealt{Ludwig1999} 
Appendix~A for a review). As hot bubbles rise, there is a 
temperature contrast between the bubble and the surroundings, and on 
sufficiently long timescales, a rising convective plume loses its heat via 
diffusion at a rate $L_\mathrm{bubble}\approx f4\pi\ell^2cP_\mathrm{rad}/\tau_\mathrm{b}$, 
where $\tau_\mathrm{b}$ is the optical depth of the 
bubble, and $f$ depends on the geometry of the bubble. The ratio of the 
heat content of the parcel to the heat lost as the parcel rises over distance $\ell$ is given by $\gamma$, the convective efficiency factor \citep[see, e.g.][]{Henyey1965,Cox1968,Ludwig1999,Kippenhahn}. In a radiation-pressure-dominated plasma ($P_\mathrm{rad}\gg P_\mathrm{gas}$), $\gamma=(\vc\tau_\mathrm{b})/(cf)$, and in a gas-pressure-dominated regime 
$\gamma=[(P_\mathrm{gas}/P_\mathrm{rad})\tau_\mathrm{b}\vc]/(2fc)$, as a 
parcel needs to evacuate the radiation field $\sim\Prad/\Pgas$ 
times in order to lose its thermal content. 
In the radiation-dominated regime then $\taut=c/\vc$, and in the gas-dominated regime $\taut=(\Prad c)/(\Pgas\vc)$. Thus up to a geometric prefactor, where $\tau_\mathrm{b}\sim\tau$, the efficiency $\gamma$ decreases with decreasing $\tau/\taut$.
In both regimes, where $\tau<\taut$, a bubble radiates a 
significant portion of its heat as it rises. 

In solar-like convection, and in evolved lower-mass stars, the transition through $\tau=\taut$ is at low enough optical depth, $\tau\sim$ a few, that it can be studied in detailed, plane parallel RHD conputations \citep[e.g.][]{Trampedach2013,Trampedach2014a,Trampedach2014,Magic2013a,Magic2013b,Magic2015,Chiavassa2018,Sonoi2019}, and incorporated into 1D stellar models via a tabulated boundary condition \citep[e.g.][]{Trampedach2014a,Salaris2015,Magic2016,Mosumgaard2018,Spada2021}.
However, in our spherical-polar near-super-Eddington RSG models, 
the large density fluctuations in the global convective plumes discussed in the previous 
section extend out into the $\tau\leq\taut$ region, and behave somewhat differently. 

\begin{figure}
\centering
\includegraphics[width=\columnwidth]{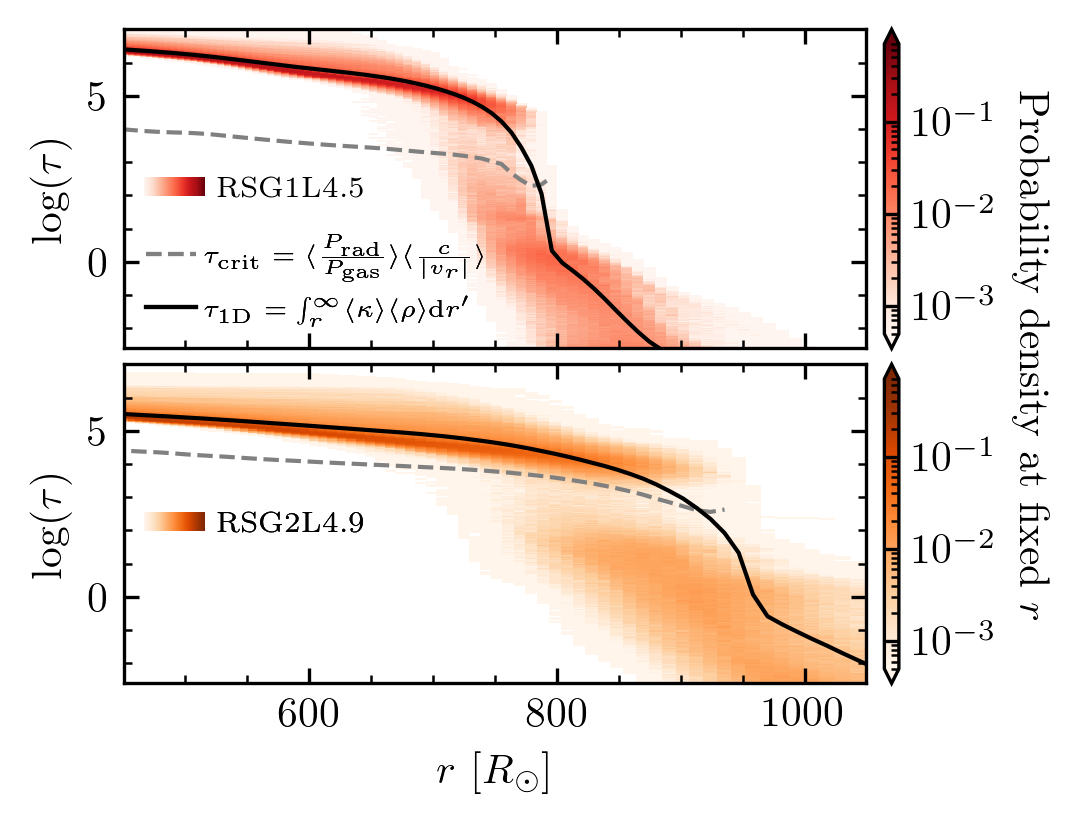}
\caption{ \label{fig:3Dtaus} Optical depth $\tau$ in characteristic snapshots 
of our RSG1L4.5 (day 4707; red, upper panel) and RSG2L4.9 (day 4927; orange, 
lower panel) models. Color saturation indicates the volume-weighted  
probability of finding a fluid element at each radial coordinate with a given $\tau$, calculated 
along radial lines of sight. Optical depth $\tau_\mathrm{1D}$, integrating the volume-averaged opacity 
and volume-averaged density, is given by the black line. The grey dashed line indicates the average 
$\taut$ at each radius (truncated outside $r=\rphot$); near the outer layers of the star, most material is either significantly 
above or below $\taut$, with little material with $\tau\approx\taut$.}
\end{figure}

Fig.~\ref{fig:3Dtaus} compares the optical depth profile integrated along 
radial lines of sight in our 3D \Athena\ simulations to the critical optical 
depth $\taut$ where we use the amplitude of the radial velocity $|\vr|=\sqrt{\vr^2}$ as our proxy for $\vc$. Due to the bimodal opacity distribution of material above and 
below H-recombination, at a given radius near where $\tau_\mathrm{1D}=\taut$, 
there is very little material with $\tau$ near $\taut$. Rather, most of the 
fluid has $\tau\gg\taut$ by over an order of magnitude, or $\tau\ll\taut$ by 
more than an order of magnitude. This is yet another signature of the 
large-scale plume structure; within a given plume, the convective velocities 
are set nonlocally, and except at interfaces between plumes there is little 
opportunity for radiative losses as fluid interacts primarily within the same plume. Comparing to the same snapshots in Fig.~\ref{fig:SLPratio}, while the entropy profiles begin to decline due to superadiabatic convection even where $\tau>\taut$ (especially in the more luminous RSG2L4.9 model), the entropy profiles decline significantly in the region where some material has $\tau<\taut$, due to the plumes losing heat via diffusion and, where $\tau\ltapprox1$, non-local radiative losses.

\begin{figure}
\centering
\includegraphics[width=\columnwidth]{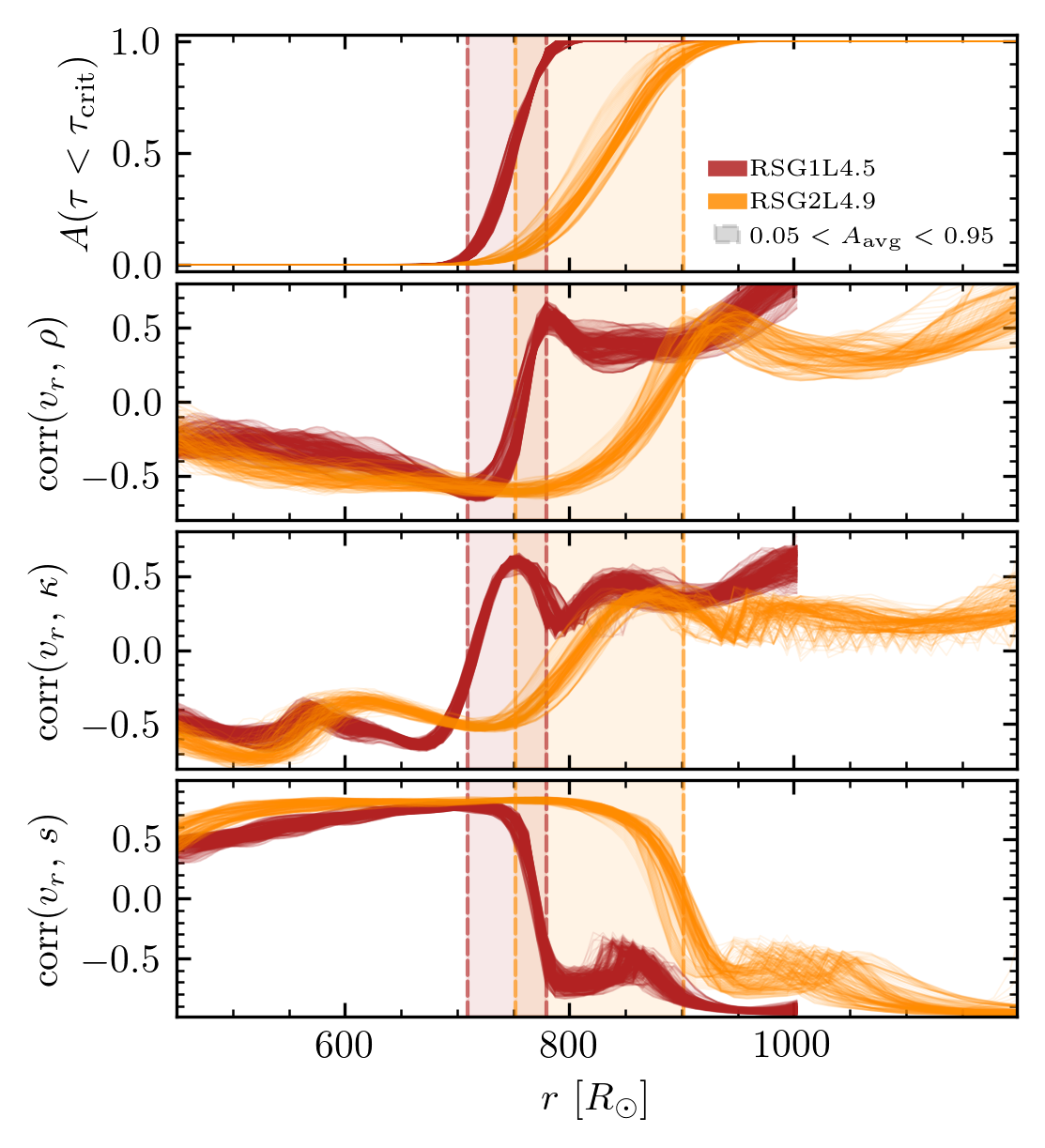}
\caption{\label{fig:3Dcorr} Fraction of the solid angle where $\tau<\taut$ along a radial line of 
sight (top panel), as well as the correlations of radial velocity with the density (second panel), 
opacity (third panel), and entropy (bottom panel). Colored lines show $\approx350$ total snapshots 
starting from day 4000 in RSG1L4.5 and day 4500 in RSG2L4.9, until the end of the simulations. The 
shaded area indicates radii where the time-average of $A(\tau<\taut)$ is between 5\% and 95\%. The 
RSG1L4.5 data are truncated at $r=1000\Rsun$, where the density approaches the density floor.}
\end{figure}

At optical depths with radiation-dominated energy transport where $\tau<\taut$, 
but still $\tau>1$, radiation forces may significantly 
impact fluid motion at high $L\sim\Ledd$. In our simulations, we observe a change in the dynamics between 
regions of high and low $\tau/\taut$. This change can be seen in 
Fig.~\ref{fig:3Dcorr}, which shows the area fraction of material with 
$\tau<\taut$, compared to correlations between the radial velocity and the 
density, opacity, and entropy of the fluid, defined by
\begin{equation}
\label{correlationfunction}
\mathrm{corr}(x,y)=\frac{\sum\left(x_{i}-\vavg{x}\right)\left(y_{i}-\vavg{y}\right)}{\sqrt{\sum\left(x_{i}-\vavg{x}\right)^{2}\sum\left(y_{i}-\vavg{y}\right)^{2}}}
\end{equation}
and the sum is taken over all zones (subscript $i$) in each radial shell. 
Where $A(\tau<\taut)=0$, the density and opacity are anti-correlated with the 
radial velocity, and the entropy is positively correlated with the radial 
velocity (where +$\vr$ is defined as moving outwards). 
This is as expected for typical MLT-like convection; the material
that sinks is denser, lower-entropy (colder), and more opaque material than the
surrounding medium. However, for the outer radii where $\tau<\taut$, the 
correlation switches and cold (low-entropy), opaque, dense regions rise! 
The shaded area indicates radii where the fluid shows a mix of $\tau>\taut$ and
$\tau<\taut$ material, quantified by where the time-average of $A(\tau<\taut)$ 
is between 5\% and 95\%, and it also captures the region in the star where the 
correlations invert. This inverted correlation is also characteristic of 
surface turbulence driven by the Fe opacity peak in younger massive stars 
(Schultz et al., in prep). In these highly luminous stars, $L$
can exceed $\Ledd$ at the $\tau=\taut$ location due to the presence
of opacity peaks, and further analysis is required to understand what
drives these near-surface dynamics. Because the nature the RHD 
turbulence changes where $\tau<\tau_\mathrm{crit}$, we thus presume for 
now that $\tau\approx\tau_\mathrm{crit}$ is an outer boundary where 
MLT treatments may cease to be appropriate in the RSG regime.

\begin{figure}
\centering
\includegraphics[width=1.05\columnwidth]{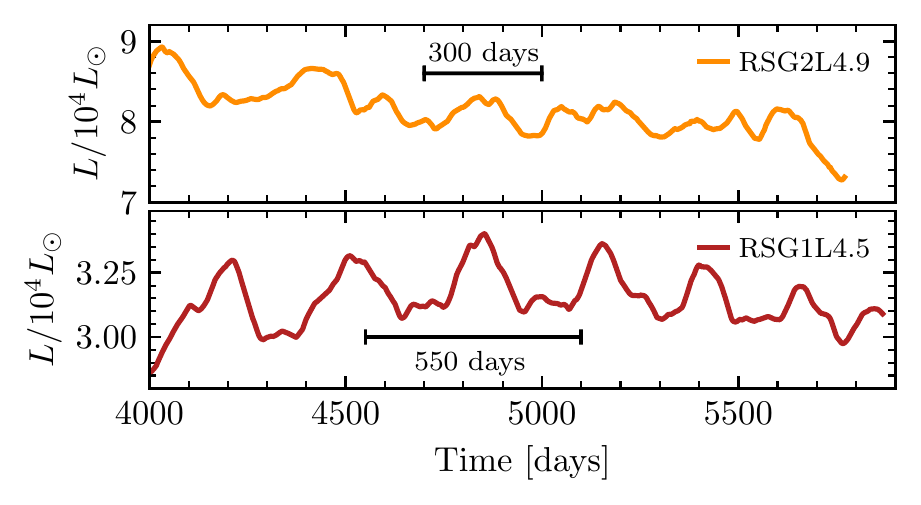}
\caption{\label{fig:lightcurves} Lightcurves starting at day 4000 for both simulation runs.  The RSG1L4.5 lightcurve is shown red on the lower panel and RSG2L4.9 in orange on the upper panel. Characteristic velocity decorrelation times for each model are indicated on the figure.} 
\end{figure}

Moreover, the observable photosphere around $\rphot$ is in this lossy, inverted-correlation, 
turbulent-pressure-dominated region! 
The convective motions here give rise to luminosity variations on timescales comparable to the timescales of the global convection cells. 
Fig.~\ref{fig:lightcurves} shows the lightcurves of the last $\approx$2000 days of our 
simulations, determined at the simulation outer boundary as $L(t)=\frac{4\pi}{\Omega_\mathrm{sim}}\int r^2F_r(t)\intd\Omega$, where $\Omega_\mathrm{sim}=\int_{\theta=\frac{\pi}{4}}^\frac{3\pi}{4}\int_{\phi=0}^\pi\intd(\cos\theta)\intd\phi$ is the solid-angle of our simulation domain. 
Fitting a second-order polynomial to the 
lightcurves\footnote{in python, using \texttt{numpy.polyfit}} and subtracting 
$L(t)-L_\mathrm{poly}(t)$, we compute the time-weighted variance as 
$\sigma_{L}^2=\sum{[L(t)-L_\mathrm{poly}(t)]^2\intdt}/\sum\intdt$, 
and the fluctuation amplitude as $\mathrm{max}[L(t)-L_\mathrm{poly}(t)]-\mathrm{min}[L(t)-L_\mathrm{poly}(t)]$. The lightcurves beyond day 4000 exhibit $\approx3\%$ mean 
luminosity fluctuations, with $\sqrt{\sigma_L^2}=0.89\times10^{3}\Lsun$ in RSG1L4.5, and $\sqrt{\sigma_L^2}=1.9\times10^{3}\Lsun$ in RSG2L4.9. 
Fluctuation amplitudes are $\approx$10\%: $3.6\times10^3\Lsun$ in RSG1L4.5 and 
$8.6\times10^3\Lsun$ in RSG2L4.9. The peak-to-peak fluctuations in the lightcurves
are irregular in time, and for RSG2L4.9 a single dominant period could not be found in the 
power spectrum.
This is likely due to the stochastic nature of the 
convective fluctuations. 
For the de-trended RSG1L4.5 power spectrum 
calculated from day 4500 onward, there is some excess 
power centered around 310 days/cycle with a 70 day spread 
resembling quasi-periodic oscillations with a wide 
window function. This flattens out when considering 
the lightcurve after day 4000, and disappears when 
considering the lightcurve from much earlier than that.
We discuss briefly in the 
Conclusions (\S\ref{sec:conclusions}) how this 
variability compares to observations.

\begin{figure*}
\centering
\includegraphics[width=\textwidth]{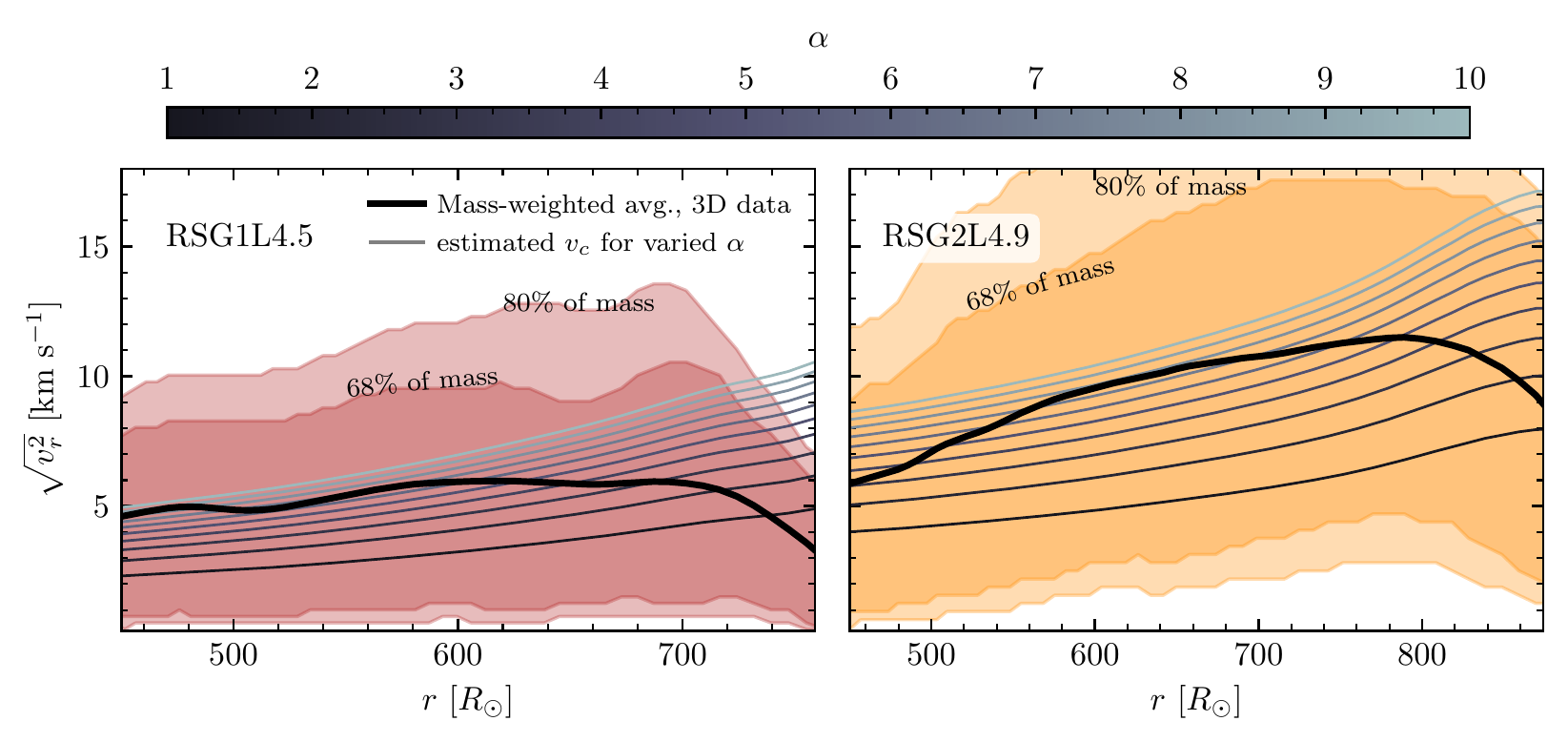}
\caption{\label{fig:alphavels} Radial fluid velocity magnitudes at characteristic snapshots of our models compared to MLT. The left panel shows RSG1L4.5 at day 4707, and the right panel shows RSG2L4.9 at day 4927. Mass-weighted average velocities are shown as thick black lines, with 68\% and 80\% of the mass lying within the dark and light shaded regions, respectively. The grey lines indicate the convective velocities predicted from MLT given the volume-averaged temperature and density profiles and the model luminosity, for integer values of $\alpha=1-10$. The plots are truncated where $\mathrm{corr}(\vr,\rho)=0$, outside of which the turbulent motions do not resemble MLT-like convection.} 
\end{figure*}

\subsection{Caveats of the 3D models}\label{sec:caveats}
There are a few caveats which we believe do not impact any of the results presented here, 
but are worth stating. First, while we include radiation pressure in the stellar interior, 
the radiation transport module in \Athena\ is not yet compatible with arbitrary gas equations of state.
As such, our assumed value of $\mu=0.645$ entails that the gas pressure may be overestimated by up to
a factor of $2$ in the outer regions with $T\ltapprox10^4$K, which could help account for the
relatively low $\Teff$ of our models. However, it should be noted that this region is exactly 
where turbulent pressure is expected to dominate over thermal pressure, which would be even more
significant if the gas pressure were lower than in our models here due to H recombination. 
Secondly, while these simulations employ full self-consistent coupling between radiation and 
hydrodynamics, the grey OPAL opacities do not account for
frequency-dependent effects. As shown by \citet{Chiavassa2011b}, non-grey opacities could lead to a 
steeper thermal gradient in the optically thin region, with weaker temperature fluctuations, which 
affects the stellar spectrum and thereby interferometric determinations of stellar radii.
The small changes in the mass within the simulation domain are dominated by IB effects 
and not outflows.
The incorporation of non-grey phenomena would also be required to place 
first-principles constraints on mass loss and other important observable stellar properties. 
Finally, while our simulation domain captures a very large fraction of the $\Omega=2\pi$ hemisphere,
the relatively few convective plumes suggest that a full $\Omega=4\pi$ simulation might yield more 
accurate cancellation of random angular momenta than our estimate, 
and may have an impact on the RSG lightcurve, which shows variability consistent with these 
stochastic convective fluctuations.

\section{Implications for 1D calculations}
\label{sec:3Dto1D}

Computational RHD models of convection enable tests of MLT assumptions, possible calibrations, and incorporation into 1D models.
A fundamental set of early 2D RHD simulations \citep{Ludwig1999} calibrated MLT parameters for 
portions of the low-optical-depth regime in $L\ll L_\mathrm{Edd}$ stars, which was followed up with 3D
simulations by \citet{Sonoi2019}, who do not definitively conclude if any particular convection 
model gives the best correspondence between 1D and 3D models, but constrain $\alpha\approx1-2$ 
across a grid of cool giant atmospheres (Red Giants with $\Teff>4000$K), in agreement with 
some observational constraints \citep[e.g.][]{Joyce2018}. Other works (e.g. 
\citealt{Trampedach2014,Magic2015,Salaris2015}) recovered similar calibrations in similar 
$L\ll\Ledd$ stars, though as convection becomes more vigorous in stars with higher luminosity and 
stronger opacity peaks and plumes take up larger and larger fractions of the star, the convective 
motions, particularly at the stellar surface, can deviate significantly from MLT 
\citep[see  e.g. discussion in][]{Trampedach2013}. 
We now discuss the implications of our 3D models for 1D calculations, focusing on the region where 
$\mathrm{corr}(\vr,\rho)<0$ so convection can be fairly compared to 
MLT's working hypothesis. Hereafter, we will refer to the location where the $\vr-\rho$ correlation inverts $(\mathrm{corr}(\vr,\rho)=0)$ as the ``correlation radius," $\Rcorr$.

\begin{figure*}
\centering
\includegraphics[width=0.48\textwidth]{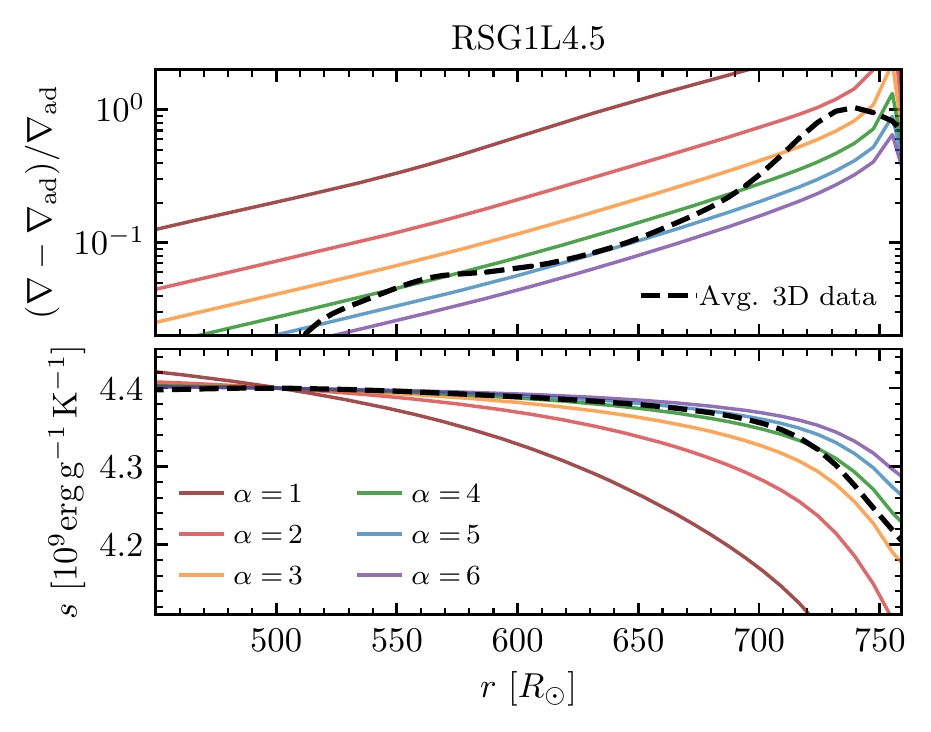}
\includegraphics[width=0.48\textwidth]{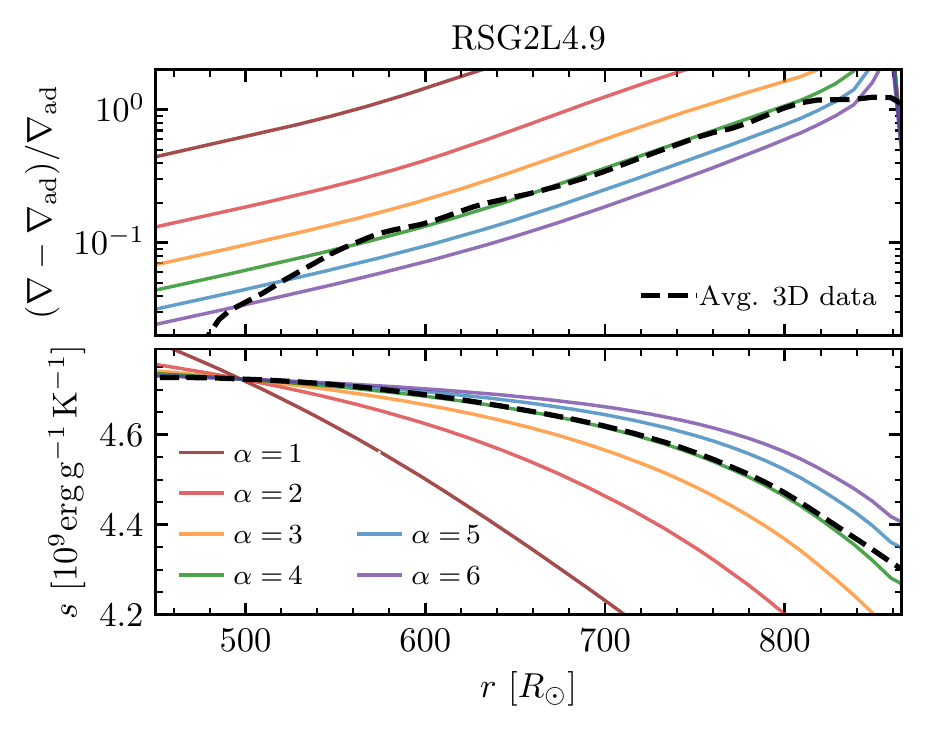}
\caption{\label{fig:MLTcal} Comparison of superadiabaticity (upper panels) and specific entropy (lower panels) derived from our 3D models (black dashed lines) and from MLT (solid colored lines), for RSG1L4.5 (left) and RSG2L4.9 (right) when $\Pturb$ is neglected. All values shown are derived from the time-averaged, shellular (volume-weighted) averaged density and temperature profiles, as well as the time-averaged luminosity at the simulation outer boundary, beyond day 4000 in RSG1L4.5 and beyond day 4500 in RSG2L4.9.} 
\end{figure*}

\subsection{Comparing Convective Velocities to MLT Expectations}

We first check the fluid velocities in our models against expectations from MLT for spherical stellar envelopes with luminosity $L$, and $\rho(r)$, and $T(r)$ profiles matching averages of our 3D models. 
Where convection carries most of the flux, as in the RSG interior, $F_\mathrm{conv}\approx L/4\pi r^2$, and from Eqs~\eqref{eq:mltflux} and \eqref{eq:mltv}, 
\begin{equation}
    \label{eq:mltvcest}
    v_{c}\approx\left(\frac{\alpha}{4}\right)^{1/3}\left(\frac{L}{4\pi r^2}{\frac{PQ}{\rho^2 c_pT}}\right)^{1/3}.
\end{equation}
Fig.~\ref{fig:alphavels} compares this expectation to the fluid motion in 
our two RSG envelope models as a function of the mixing length parameter 
$\alpha$, with the diagnostic velocity taken to be $\sqrt{\vr^2}$ in the 3D 
models.
We represent the 3D data via bands, with 80\% of the mass having velocities lying within the light 
colored regions, and 68\% having velocities within the darker colored 
regions. The mass-weighted averages are indicated by the thick black 
lines. For clarity, we show here the comparison for individual model 
snapshots; the time-averaged profiles display similar behavior. The azimuthal and polar velocity profiles are comparable, with $\langle\sqrt{v_\theta^2}\rangle_m\approx\langle\sqrt{v_\phi^2}\rangle_m\approx5-8$ km/s in RSG1L4.5 and 7-9 km/s in RSG2L4.9, with large scatter ($\gtapprox\pm5$km/s), and radial motion accounts for $\approx1/3-1/2$ of the turbulent kinetic energy density.
We see good (factor of $\approx2$) agreement between the convective 
velocities predicted by MLT and the 3D models, and the scatter in 
convective velocities is much larger than the factor of $10^{1/3}$ 
introduced by varying $\alpha$ by a factor of 10.  
In both models, the velocity profile is flatter across a larger radial
domain than MLT would predict for a fixed $\alpha$.
We speculate that this can be attributed to the nonlocal, large-scale nature of the plumes, 
as the velocity profile is set by the motion of a mixture of plumes which do not change significantly 
over the simulation domain; this is also noted in, e.g., \citet{Brun2009} in 3D simulations of RGB stars.

\subsection{Calibration of Mixing Length Parameters in the Absence of $\Pturb$}
\label{sec:alphacal1}
Convective efficiency is important in determining the stellar radius as 
discussed in detail in \S\ref{sec:background}; therefore it is
valuable to have a first-principles calibration of mixing length 
parameters, especially $\alpha$, within the RSG regime
motivated by 3D models. Because the nature of the turbulent energy and momentum transport changes outside $\Rcorr$, we treat $\Rcorr$ as an outer boundary beyond which MLT treatments cannot be calibrated, and perhaps cease to be appropriate, in the high-luminosity RSG regime.

Most 1D stellar-evolutionary models do not account for 
turbulent pressure, and when included, it is a challenge (see discussion in \citealt{Trampedach2014}), so we first explore the case where $P=\Ptherm=\Prad+\Pgas$.
We generate a 1D model from the 3D simulations by finding the time-averaged, 
volume-averaged radial density and temperature profiles from each 3D 
simulation run ($\rho_\mathrm{1D}(r)$ and $T_\mathrm{1D}(r)$, respectively). 
We choose volume-averages along surfaces of constant gravity (radial coordinates) due to the 
loosely-bound nature of the envelope, though where $r<\Rcorr$ different averages do not significantly affect our results.
We calculate $\kappa$ from these profiles using the OPAL tables. 
The total luminosity is taken to be the time-averaged 
luminosity in the outermost zone $L=\Lsurf$, up to the end of the 
simulation starting from day 4000 in RSG1L4.5 and from day 4500 in RSG2L4.9. We assume an EOS 
of ideal gas + radiation with $\mu=0.645$, as in our 3D model, which is appropriate for 
$r<\Rcorr$ as $T\gtapprox10^4$K. We then solve the \citet{Henyey1965} MLT equations assuming $y=3/4\pi^2$, and 
consider only material inside $\Rcorr$ (where $\tau>\taut\gg1$) for different values of $\alpha$ 
(see Appendix~\ref{sec:AppendixA} for more specific details).

The upper panels of Fig.~\ref{fig:MLTcal} show the comparison between the 
superadiabaticity, expressed as $(\grad-\gradad)/\gradad$, using $\grad$ predicted by MLT and 
$\grad$ derived directly from the averaged 3D data. The x-axis limits are $450R_\odot$ and 
$\Rcorr$, respectively. We see significant deviations between $\grad$ from $\gradad$, with 
nearly-adiabatic behavior in the interior and increasing superadiabaticity outward. The lower 
panels show entropy profiles, which are often used to calibrate MLT parameters to 3D atmosphere
models in more compact, less luminous stellar environments \citep[e.g.][]{Trampedach2014,Magic2015,Magic2016,Sonoi2019}. For our 
averaged 3D data, we calculate $s$ including radiation and gas entropy, $s=\frac{k_\mathrm{B}}{\mu m_p}\ln\left(T^{3/2}/\rho\right)+\frac{4}{3}a_rT^{3}/\rho$, where $T$ is in K and $\rho$ is in g/cm$^3$,
and for MLT we integrate $\intd s=c_P\,\intd\ln T\left[1-\gradad/\grad\right]$ 
using $\grad$ given by MLT, connecting to the nearly-adiabatic $r=500\Rsun$ location to ensure
agreement in the additive constant.
Both models display nice agreement with $\alpha=4$ in the interior. At larger radii, the RSG1L4.5 model (left panels) exhibits greater superadiabaticity than implied by $\alpha=4$, in better agreement with $\alpha=2-3$. This contributes to the entropy profile, which falls more steeply than $\alpha=4$ and approaches the value predicted by $\alpha=3$ in our region of consideration. The more luminous RSG2L4.9 model (right panels of Fig.~\ref{fig:MLTcal}) closely follows the $\alpha=4$ predictions throughout most of the domain of interest, with generally excellent agreement for the entropy profile, becoming more shallow as $r$ approaches $\Rcorr$.

\subsection{Estimating $\Pturb$ in a 1D model and MLT Implications}\label{sec:PturbMLT}

\begin{figure}
\centering
\includegraphics[width=\columnwidth]{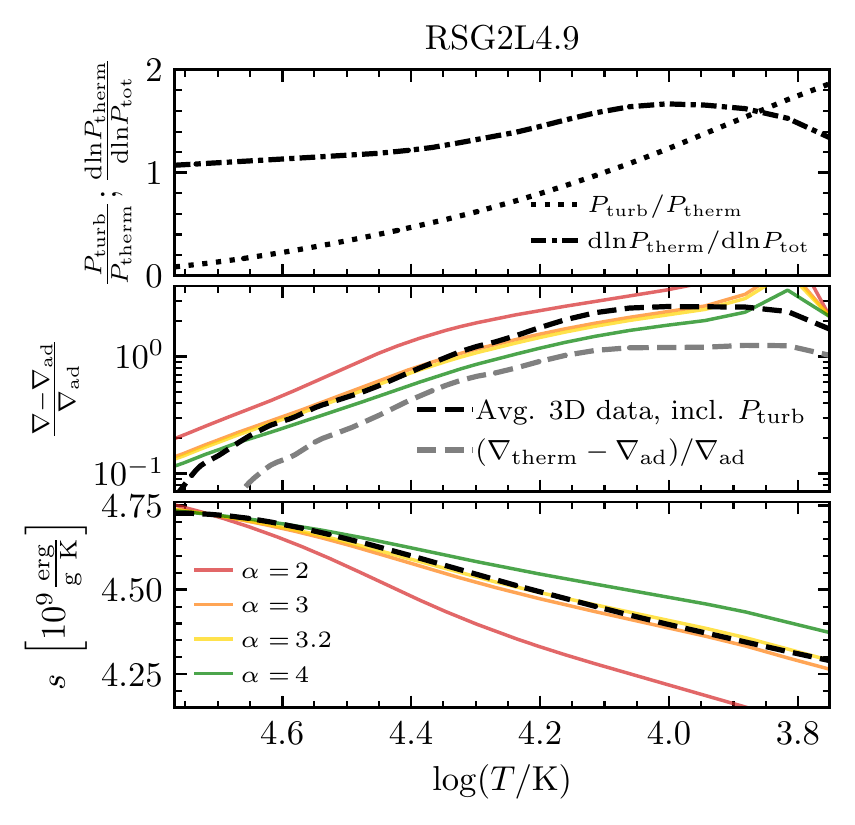}
\caption{\label{fig:MLTcalPturb} Impact of turbulent pressure on mixing length calibration.  Comparison of superadiabaticity (middle panel) and entropy 
(lower panel) are shown, for our averaged 3D RSG2L4.9 model (black dashed lines) and MLT with 
different $\alpha$ (solid colored lines) as a function of temperature. 
The upper panel shows $\Pturb/\Ptherm$ (dotted line) and $\intd\ln\Ptherm/\intd\ln\Ptot$ (dash-dot line), which are included 
in the MLT calculation and in the averaged 3D $\grad$. The grey dashed line in  the middle panel
shows $\grad_\mathrm{therm}=\intd\ln T/\intd\ln\Ptherm$ to facilitate direct comparison to 
Fig.~\ref{fig:MLTcal}. All values shown are derived from the time-averaged (beyond day 4500), 
shellular (volume-weighted) averaged density, temperature, and kinetic energy profiles, as well
as the time-averaged luminosity at the simulation outer boundary. The x-axis extends from $r=450\Rsun$ to $\Rcorr=865\Rsun$.} 
\end{figure}

In a vigorously convective stellar envelope, turbulent pressure can become comparable to the 
thermal pressure and provide hydrostatic support. A fully self-consistent 1D implementation of 
turbulent pressure in 1D models remains an open challenge, as the inclusion of turbulent 
pressure leads to unrealistically steep pressure gradients near convective boundaries, 
especially near the stellar surface \citep{Trampedach2014}. 
In MLT, turbulent pressure can be incorporated by modifying the pressure scale height and the adiabatic temperature gradient. Using the chain rule to include 
$\Ptot=\Pturb+\Ptherm$, the modified adiabatic temperature gradient, $\nabla_\mathrm{ad}'$, is given by \citep{Henyey1965},
\begin{equation}
    \nabla_\mathrm{ad}'=\left(\frac{\intd\ln{T}}{\intd\ln{\Ptherm}}\right)_\mathrm{ad}\times\frac{\intd\ln{\Ptherm}}{\intd\ln{P_\mathrm{tot}}}.
\end{equation}
The substitution $\gradad\rightarrow\gradad^\prime$ is then made where $\gradad$ appears in the
MLT equations (\citealt{Henyey1965}; see also our Appendix~\ref{sec:AppendixB}) and $H$ is calculated as $\Ptot/
\rho g$.
The lack of a reliable method to estimate $P_\mathrm{turb}$ inhibits such an incorporation in most 1D MLT 
implementations. For convenience, definitions of different gradients we used are also summarized in Appendix~\ref{sec:AppendixB}.

Quantifying the pressure associated with turbulent kinetic energy densities from 3D RHD models allows us to explore how the 1D gradients are modified for these stars. The nonlocal nature of the convective motions means that the 
characteristic fluid velocity used in calculating $\Pturb$ is not simply identified 
with the velocity parameter in MLT. Therefore, in order to estimate the impact of turbulent 
pressure on the thermodynamic gradients and recovered values of $\alpha$, we determine 
$\intd\ln\Ptherm/\intd\ln\Ptot$ directly from 1D averages of our 3D models. 
For this initial exploration, we calculate $\Pturb$ and thereby
$\intd\ln\Ptherm/\intd\ln\Ptot$ using the time-average of the angle-averaged $\Pturb=\langle\rho\vr\vr\rangle$. We then use \citet{Henyey1965}'s formula with 
turbulent pressure motivated by the 3D data to solve for $\grad$ at different values of $\alpha$.

Fig.~\ref{fig:MLTcalPturb} shows the results of this exercise for the RSG2L4.9 model. 
The upper
panel shows the adiabatic correction term ($\intd\ln\Ptherm/\intd\ln\Ptot$; dash-dot line), as well as the ratio of turbulent pressure to thermal pressure (dotted line). 
The value of $\grad=\intd\ln T/\intd\ln\Ptot$ from the averaged 3D data, for which we now include turbulent pressure as $\Ptot=\Ptherm+\Pturb$,
is shown by the black dashed line in the middle panel. 
For direct comparison to Fig.~\ref{fig:MLTcal}, we compare $\nabla$ here to $\gradad$ rather than $\gradad^\prime$.
To further facilitate direct comparison, the grey dashed line in the middle panel shows the 
value of $\gradth=\intd\ln{T}/\intd\ln{\Ptherm}$, which was taken to be equivalent to 
the true $\grad$ in \S\ref{sec:alphacal1} and is equivalent to the dashed black line in the upper right panel of Fig.~\ref{fig:MLTcal}. The lower panel shows the entropy, 
calculated using $\gradad^\prime$ and $\grad$.
The black dashed line in the lower panel gives the entropy profile for our 3D-motivated 1D model, which is equivalent to the 
black dashed line in the lower right panel in Fig.~\ref{fig:MLTcal}, as the turbulent pressure terms 
cancel in the expression for $s$ (i.e. $\gradat/\grad=\gradad/\gradth$). For the MLT values, shown by the colored lines, each value of 
$\alpha$ recovers a larger value of $\nabla$ compared to \S\ref{sec:alphacal1},
but a slightly shallower $s$ profile (as the turbulent pressure correction in $\gradad^\prime$ 
accounts for a greater portion of the $\grad-\gradad$ excess). Therefore, if a 1D stellar 
evolution code were to include a turbulent pressure correction to MLT using the 
\citet{Henyey1965} parameters, we would recommend a value of $\alpha=3.2$ from this model.

\section{Discussion \& Conclusions}\label{sec:conclusions}

We have constructed global 3D radiation hydrodynamical simulations in the RSG regime which include an 
accurate gravitational potential and radiation pressure in the convective interior for the first time.
These simulations span $\approx$70\% of the $2\pi$ 
hemisphere and yield predictions for the turbulent structure and dynamics from the middle of the 
convective envelope out beyond the photosphere. 
Our incorporation of radiation pressure in optically thick regions has enabled realization 
of the expected nearly-constant entropy profile and convective-luminosity domination
in the convective interior. 
In agreement with \citet{Freytag2002} and \citet{Chiavassa2009} we find that the convection is dominated by a few large-scale 
plumes which flow through most of the simulation domain and survive for 
timescales of $\approx$300 and $\approx$550 days (for RSG1L4.5 and RSG2L4.9, respectively; see 
Fig.~\ref{fig:sscorr}). 
When the models reach a convective steady state, RSG1L4.5 has 
$\log(L/\Lsun)\approx4.5$ and $\rphot\approx800\Rsun$, and RSG2L4.9 has $\log(L/\Lsun)\approx4.9$ and 
$\rphot\approx900\Rsun$. 

Both models display $\approx$10\% variation in luminosity owing to the 
large-scale turbulent surface structure (see Fig.~\ref{fig:lightcurves}). 
Temporal observations (see, e.g. 
\citealt{Kiss2006, Soraisam2018, Conroy2018, Chatys2019, Ren2019, Soraisam2020}) reveal RSG 
variability on timescales of a few hundred to thousands of days in a variety of host environments. These signals include both periodic and stochastic behavior, 
with increasing ubiquity of larger-amplitude fluctuations for brighter stars. 
In M31, for example, all RSGs brighter than $\log(L/\Lsun)>4.8$ display lightcurve fluctuations with
 $\Delta m_R>0.05\ \mathrm{mag}$, up to around $\Delta m_R\approx0.4$ \citep{Soraisam2018}.
Periodic variability is interpreted as radial pulsations 
\citep{Stothers1969,Stothers1971,Guo2002}, likely driven by a hydrogen 
ionization region inside the convective envelope \citep{Heger1997, Yoon2010}. The
stochastic fluctuations \citep[e.g.][]{Ren2020} qualitatively agree with our models, and we intend 
further analysis to compare these convective models directly to observations.

In the outer stellar layers, radiation carries an increasing fraction of the total luminosity as 
convection becomes lossy. This transition is associated with reaching an optical depth $\tau<\taut\approx100$. 
Moreover, large density fluctuations and appreciable bimodality in $\kappa$ and $T$ lead to a range of radii with increasing amounts of material at $\tau<\taut$ (see Figs.~\ref{fig:3Dpanels},\ref{fig:3Dtaus}).
In the region where $\tau$ along some lines of sight falls below $\taut$, the 
correlations of radial velocity with the fluid density, entropy, and opacity fluctuations invert from what is 
characteristic of convective fluid motions (see Fig.~\ref{fig:3Dcorr}); indeed the denser, lower-entropy, higher-opacity material rises!
These inverse correlations at $A(\tau<\taut)<1$ where $L$ locally exceeds $\Ledd$ will not be seen if radiation pressure is not inlcuded. 
The change in the nature of convective motions in the outermost stellar layers of these 
highly luminous RSGs also prohibits a comparison to MLT treatments. Hence, we define the radius where these correlations invert
as $\Rcorr$, taking it as an outer boundary where MLT-like treatments cease to be appropriate. 

Inside $\Rcorr$, where MLT is applicable, the velocity profiles are flatter than MLT-like convection 
due to the nonlocal, large-scale convective plumes, but display good order-of-magnitude agreement (see 
Fig.~\ref{fig:alphavels}). By comparing entropy profiles and superadiabatic
gradients inside $R<\Rcorr$, we find from our 3D simulations that the mixing length $\alpha$ 
appropriate for convection in this regime is $\alpha\approx3-4$ (see Fig.~\ref{fig:MLTcal} for models 
which neglect pressure from the turbulent motions and Fig.~\ref{fig:MLTcalPturb} which includes an estimate for such a correction). This convective efficiency is more consistent with estimates of larger-than-solar mixing lengths from the HR position of RSG populations \citep{Chun2018}, supernova color evolution \citep{Dessart2013}, and even some 3D treatments of the Sun which compare conventional MLT to other prescriptions for handling the different flux terms \citep[e.g.][]{Porter2000}.
Future work of immediate interest will focus on better understanding the nature and implications 
of the surface turbulence outside of $\Rcorr$. Similar inverted-correlation 
behavior is also seen in other simulations of luminous stars (e.g. in OB-star envelopes; \citet{Schultz2022}), but not in simulations of solar-like convection (e.g. \citealt{Stein1998}),
and may owe to RHD effects where $\taut\gg1$ and $L\gtapprox\Ledd$.

In addition to exhibiting large density fluctuations which increase at large radii, 
the \Athena\ RSG models display shallower density profiles in their outer stellar 
halos compared to traditional 1D hydrostatic models, and material near $\tau=1$ 
($\approx$50$-$100$\Rsun$ beyond $\Rcorr$) reaches densities 1\hyphen2 orders of magnitude lower than 
barren 1D model photospheres. In the eventual explosion of a RSG as a Type IIP Supernova, shock 
propagation (and therefore the SN emission) may be moderated by these 3D envelopes. Early SN emission 
(first $\approx30$ days) is sensitive to the outermost $<0.01-0.1\Msun$ of material; thus the 
inverted-correlation surface-turbulent outer halo defines the emitting region for the shock breakout 
and shock cooling phases of SN evolution. These phases have been studied
extensively for 1D hydrostatic models with a well-defined outer radius (e.g., 
\citealt{Nakar2010,Morozova16,Shussman2016,Sapir2017,Faran2019,Kozyreva2020}), 
but not for fundamentally 3D envelopes.
The outer halo of material will also modify the predicted UV shock breakout signatures.
The extent to which the 3D envelope properties discussed above may aid in our understanding of
early-time Type IIP SN emission is thus an exciting avenue for our future exploration.


\acknowledgements
We thank the anonymous referee for detailed discussion and comments which have improved this manuscript.
We would like to thank William Schultz and Tin Long Sunny Wong for scientific and aesthetic feedback, 
and Andrea Antoni, Matteo Cantiello, Eliot Quataert, and Benny Tsang for invaluable correspondences. We 
also especially thank Bill Paxton and Josiah Schwab for their continued support and advancement of \MESA's
capabilities, and for valuable discussions along the way.

J.A.G. acknowledges the National Science Foundation (NSF) GRFP grant No. 1650114. This research was 
supported by the NSF under grants ACI-1663688 and PHY-1748958, and by the NASA ATP grant 
ATP-80NSSC18K0560. This research also benefited from interaction that were supported by the Gordon 
and Betty Moore Foundation through Grant GBMF5076. The Flatiron Institute is supported
by the Simons Foundation.
Resources supporting this work were provided by the NASA High-End Computing (HEC) program 
through the NASA Advanced Supercomputing (NAS) Division at Ames Research Center.
We acknowledge support from the Center for Scientific Computing from the CNSI, MRL: an NSF MRSEC (DMR-1720256) and NSF CNS-1725797.
This research made extensive use of the SAO/NASA Astrophysics Data System (ADS).

\software{Computational models utilized \MESA\ \citep{Paxton2011,Paxton2013,Paxton2015,Paxton2018,Paxton2019} and \Athena\ \citep{Stone2020}. Analysis
made significant use of the following packages:
\texttt{py\_mesa\_reader} \citep{MesaReader}, 
\texttt{NumPy} \citep{Numpy}, \texttt{SciPy} \citep{SciPy}, and
\texttt{matplotlib} \citep{Matplotlib}. Figure colors made use of the additional \texttt{python} package \texttt{cmocean} \citep{cmocean}.
}

\appendix
\section{MLT Calibration Details and Sensitivities}
\label{sec:AppendixA}

In MLT, as deployed by \citet{Henyey1965}, the optical 
thickness of a bubble is $\omega=\kappa\rho\ell$, akin to $\tau_\mathrm{b}$ discussed in \S\ref{sec:fund3D}, which is typically comparable to the optical depth to the surface ($\tau$) when the opacity is not changing drastically. 
The convective efficiency parameter is then given by 
\begin{equation}\label{eq:mltgamma}
\gamma=\frac{\grad-\grade}{\grade-\gradad}=\gamma_0\vc
\end{equation}
where $\gamma_0=c_{p}\rho/\left(8\sigma_\mathrm{SB}T^{3}\theta\right)$, 
$\theta=\omega/(1+y\omega^2)$, and $y$ depends on the geometry of the bubble. 
We solve for $\gamma$ via the cubic equation 
\begin{equation}
    \label{eq:mltcubic}
    \gamma+\gamma^{2}+\phi\gamma^{3}=\frac{gHQ\left(\alpha\gamma_{0}^{2}\right)}{\nu}\left(f \nabla_{\mathrm{rad}}-\nabla_{\mathrm{ad}}\right),
\end{equation}
where $F=\Lsurf/4\pi r^2=16\sigma T^{4}\nabla_{\mathrm{rad}}/3\kappa\rho H$ defines $\nabla_\mathrm{rad}$ as the gradient required to carry all flux by radiative diffusion, $\phi=\frac{3}{4}f\omega\theta$, $\nu=8$, and $f=1$ as $\tau>\taut\gg1$ inside $\Rcorr$. 
For an ideal gas + radiation, EOS properties vary with $\alpha_P\equiv\Prad/\Pgas$ (\citealt{Mihalas1984}; $P$ subscript added to distinguish from $\alpha=\ell/H$), with 
\begin{equation} \label{eq:cpeos}
c_P=\frac{5}{2}\frac{k_B}{\mu m_p}\left(1+8\alpha_P+\frac{32}{5}\alpha_P^2\right),
\end{equation}
and
\begin{equation}\label{eq:gradadeos}
\gradad=\frac{1+5\alpha_P+4{\alpha_P}^2}{\frac{5}{2}+20\alpha_P+16{\alpha_P}^2}.
\end{equation}

From this, MLT yields a prediction for $\nabla$, which we compare to the gradients derived from $\rho_\mathrm{1D}$ and $T_\mathrm{1D}$:
\begin{equation}
\label{eq:mltgrad}
\nabla=\frac{(1+\gamma)f\nabla_{\mathrm{rad}}+\phi\gamma^{2}\nabla_{\mathrm{ad}}}{1+\gamma+\phi\gamma^{2}}.
\end{equation} 

\begin{figure*}
\centering
\includegraphics[width=0.48\textwidth]{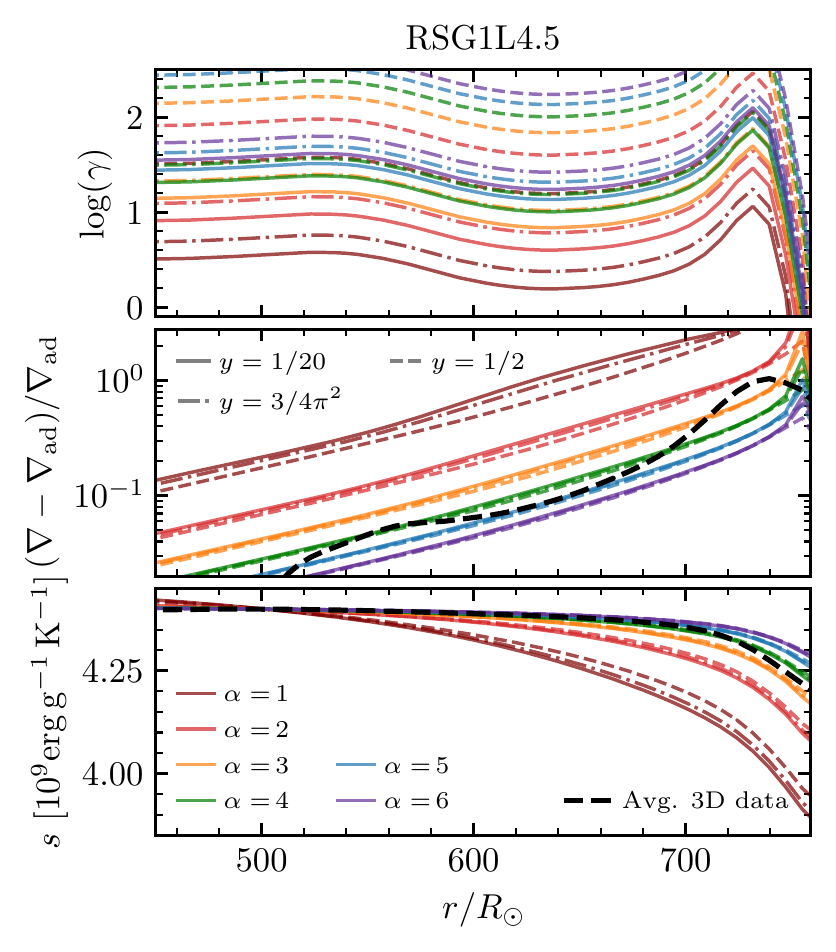}
\includegraphics[width=0.48\textwidth]{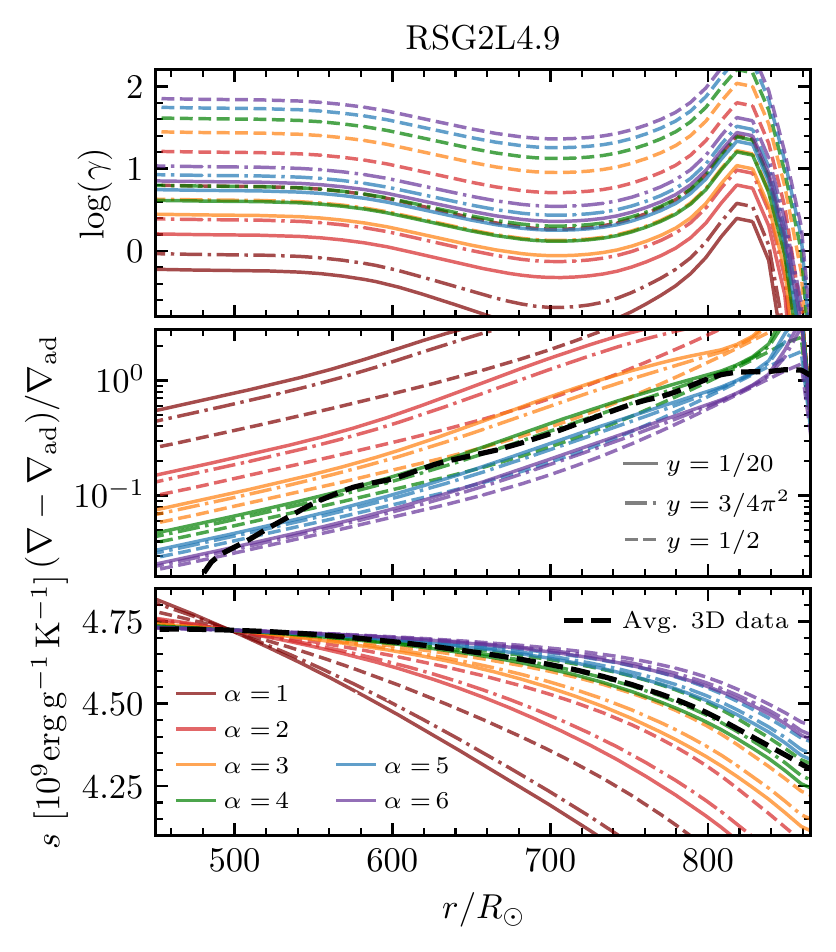}
\caption{\label{fig:apxfig1} Comparison of efficiency factor $\gamma$ (upper panels),  superadiabaticity (middle panels) and entropy (lower panels) derived from our 3D models (black dashed lines) and from MLT (pastel colored lines) against profiles derived from RSG1L4.5 (left) and RSG2L4.9 (right) when $\Pturb$ is neglected. Each color corresponds to a different value of $\alpha$, and each linestyle (solid, dash-dot, dashed) corresponds to a different value of $y$.} 
\end{figure*}

Following \citet{Henyey1965}, we use $y=3/4\pi^2$ for our analysis in \S\ref{sec:3Dto1D}.
We repeated this analysis varying $y$ for values ranging from $1/20$, which is the prediction 
for a parabolic temperature distribution inside a bubble, to $1/2$ (as used by 
\citet{BohmVitense1958}) which corresponds to a linear temperature distribution \citep[see discussion in][]{Henyey1965}. This is shown in Fig.~\ref{fig:apxfig1}.
The region inside $r<\Rcorr$ is in the limit of higher $\omega$ ($\tau\gg1$),
so $\gamma_0\appropto y$, leading to a strong $y$-dependence in $\gamma$ for both models. 
However, variations in $\gamma$ lead to large differences in $\grad$ and $s$ only when 
$\gamma\ltapprox1$. For the RSG1L4.5-derived model, $\omega$ is sufficiently large due to the 
slightly larger envelope mass and smaller radius, so fractional changes in $\gamma$ do not 
lead to significant differences in $\nabla$ or the recovered entropy profile except for the
$\alpha=1$ line (which disagrees with the model profiles). 
In the case of RSG2L4.9, $\omega$ is smaller due to the lower envelope density, so changes 
in $y$ do affect the recovered superadiabatic gradient and entropy profiles even for 
$\alpha\ltapprox3$, with higher values of $y$ leading to smaller $\grad-\gradad$ and flatter 
$s$ profiles. However, this effect is still not substantial for $\alpha=4$, which also agrees 
best with the model. In all cases, the variation in $\nabla$ and $s$ introduced by varying 
$y$ is dominated by differences with different $\alpha$. 

Comparing the luminosity carried by radiation recovered by MLT to the time-averaged 
shell-averaged $F_r$ of the 3D models, there is good agreement between the MLT values 
in both models within $\approx$5\% for $r\ltapprox700\Rsun$. Outside of those locations,
however, MLT predicts dramatically lower radiative fluxes and higher convective fluxes due to 
the presence of the H opacity peak. This is not surprising for two reasons. First, we 
consider $\kappa$ from a 1D OPAL call, where the H opacity spike is sharper (see the bottom 
panels of Fig.~\ref{fig:3Dpanels}) compared to the 3D data which displays a bimodal 
distribution of $\kappa$ in a given radial shell. Secondly, different values of $\tau$ along different
lines of sight where there is appreciable bimodality (see Figs.~\ref{fig:3Dpanels},\ref{fig:3Dtaus}) 
allow radiation to carry more of the flux than one would expect from radiative 
diffusion through a 1D shell with no density fluctuations. 

\section{Gradient Definitions with and without Turbulent Pressure}
\label{sec:AppendixB}
For convenience, we state here how the above equations include the \citet{Henyey1965} turbulent-pressure 
correction. When $\Ptot=\Ptherm+\Pturb$ is included, the modified Eq.~\ref{eq:mltgamma} is
\begin{equation}\label{eq:mltgammapturb}
\gamma=\frac{\grad-\grade}{\grade-\gradat}=\gamma_0\vc,
\end{equation}
Eq.~\ref{eq:mltcubic} becomes 
\begin{equation}
    \label{eq:mltcubicpturb}
    \gamma+\gamma^{2}+\phi\gamma^{3}=\frac{gHQ\left(\alpha\gamma_{0}^{2}\right)}{\nu}\left(f \nabla_{\mathrm{rad}}-\gradat\right),
\end{equation}
and Eq.~\ref{eq:mltgrad} becomes 
\begin{equation}
\label{eq:mltgradpturb}
\nabla=\frac{(1+\gamma)f\nabla_{\mathrm{rad}}+\phi\gamma^{2}\gradat}{1+\gamma+\phi\gamma^{2}}.
\end{equation} 

For clarity, the definitions of the relevant gradients are given in Table~\ref{tab:grads} (on the next page).
\begin{table}
\begin{center}
\begin{tabular}{ | c || c |}
 \hline
 gradient & Definition  \\ \hline\hline
$\nabla$& actual $\frac{\intd\ln T}{{\intd\ln P}}$ in the star \\  \hline
$\gradth$& $\frac{\intd\ln T}{{\intd\ln\Ptherm}}$ in the star \\  \hline
$\grade$& $\nabla$ inside an eddy as it moves\\  \hline
$\gradad$& $\left(\frac{\intd\ln T}{{\intd\ln\Ptherm}}\right)_\mathrm{ad}$ from the fluid properties \\  \hline
$\gradat$& $\gradad\times \frac{d\ln{P_\mathrm{th}}}{d\ln{P_\mathrm{tot}}}$\\  \hline
$\grad_\mathrm{rad}$& $\grad$ required to carry $\Lsurf$ solely by radiative diffusion$=(3\Lsurf\kappa\rho H)/(64\pi r^2\sigma_\mathrm{SB}T^{4})$
\\  \hline
\end{tabular}
\end{center}
\caption{Definitions of various gradients discussed in this work.\label{tab:grads}}
\end{table}

\bibliographystyle{aasjournal}
\singlespace
\typeout{} 
\bibliography{RSG3D.bib}

\end{CJK*}
\end{document}